\documentclass[twocolumn,prd,superscriptaddress,altaffilletter,nofootinbib]{revtex4-1}
\usepackage{amsmath}
\usepackage{amssymb}
\usepackage{amsfonts}
\usepackage{comment}
\usepackage{graphicx}
\usepackage[colorlinks=true,citecolor=blue,urlcolor=blue]{hyperref}
\usepackage{acronym}
\usepackage[usenames,dvipsnames]{xcolor}
\usepackage{xspace}
\usepackage{tikz}
\usepackage{siunitx}
\usepackage{makecell}
\usepackage{multirow}
\usepackage{bm}
\usepackage{romannum}
\usepackage{xcolor}
\usepackage{orcidlink}
\usepackage[caption=false]{subfig}
\usepackage{ulem}

\usetikzlibrary{arrows,shapes,trees,decorations.pathreplacing}
\allowdisplaybreaks

\newcommand{\mb}{m_{\mathrm{b}}}   
\newcommand{\fb}{f_{\mathrm{b}}}   
\newcommand{\fobs}{f_{\mathrm{obs}}}   
\newcommand{\tauc}{\tau_{\mathrm{c}}}   
\newcommand{\vvir}{v_{\mathrm{vir}}}   
\newcommand{\pSHM}{p_{\mathrm{SHM}}}   
\newcommand{\ds}{\rho}   
\newcommand{\dsinc}{\rho_{\mathrm{inc}}}
\newcommand{\dcov}{\Sigma}
\newcommand{\ncov}{\mathcal{N}}
\newcommand{\scov}{\mathcal{H}}
\newcommand{\Ttot}{T_{\mathrm{tot}}}

\newcommand{\fap}{p_{\mathrm{FA}}}
\newcommand{\cl}{p_{\mathrm{CL}}}
\newcommand{\incoherent}{incoherent sum\xspace}
\newcommand{\incoherenth}{incoherent--sum\xspace}
\newcommand{\coherent}{coherent SNR\xspace}
\newcommand{\Coherent}{Coherent SNR\xspace}

\newcommand{\e}{\mathrm{e}}
\newcommand{\iu}{\mathrm{i}}

\definecolor{forestgreen}{HTML}{228B22}

\newcommand{\ulbdm}[0]{\ac{ULBDM}\xspace}
\newcommand{\gw}[0]{\ac{GW}\xspace}
\newcommand{\gws}[0]{\acp{GW}\xspace}
\newcommand{\dm}[0]{\ac{DM}\xspace}
\newcommand{\snr}[0]{\ac{SNR}\xspace}
\newcommand{\psd}[0]{\ac{PSD}\xspace}

\begin{document}
\begin{flushleft}
{\rm RESCEU-15/24}
\end{flushleft}

\title{Enhancing the sensitivity to ultralight bosonic dark matter using signal correlations}

\author{Soichiro Morisaki \orcidlink{0000-0002-8445-6747}}
\affiliation{Institute for Cosmic Ray Research, The University of Tokyo, 5-1-5 Kashiwanoha, Kashiwa, Chiba 277-8582, Japan}
\author{Jun'ya Kume \orcidlink{0000-0003-3126-5100}}
\affiliation{Dipartimento di Fisica e Astronomia ``G. Galilei'', Universit\`a degli Studi di Padova, via Marzolo 8, I-35131 Padova, Italy}
\affiliation{INFN, Sezione di Padova, via Marzolo 8, I-35131 Padova, Italy}
\affiliation{Research Center for the Early Universe (RESCEU), Graduate School of Science, The University of Tokyo, 7-3-1 Hongo, Bunkyo-ku, Tokyo 113-0033, Japan}
\author{Takumi Fujimori \orcidlink{0009-0001-4150-6273}} 
\affiliation{Department of Physics, Graduate School of Science, Osaka Metropolitan University, 3-3-138 Sugimoto-cho, Sumiyoshi-ku, Osaka City, Osaka 558-8585, Japan}
\author{Yuta Michimura \orcidlink{0000-0002-2218-4002}}
\affiliation{Research Center for the Early Universe (RESCEU), Graduate School of Science, The University of Tokyo, 7-3-1 Hongo, Bunkyo-ku, Tokyo 113-0033, Japan}
\affiliation{Kavli Institute for the Physics and Mathematics of the Universe (Kavli IPMU), WPI, UTIAS, The University of Tokyo, 5-1-5 Kashiwanoha, Kashiwa, Chiba 277-8568, Japan}

\begin{abstract}

In recent years, numerous experiments have been proposed and conducted to search for \ulbdm.
Signals from \ulbdm in such experiments are characterized by extremely narrow spectral widths.
A near-optimal detection strategy is to divide the data based on the signal coherence time and sum the power across these segments.
However, the signal coherence time can extend beyond a day, making it challenging to construct contiguous segments of such a duration due to detector instabilities.
In this work, we present a novel detection statistic that can coherently extract \ulbdm signals from segments of arbitrary durations.
Our detection statistic, which we refer to as \coherent, is a weighed sum of data correlations, whose weights are determined by the expected signal correlations.
We demonstrate that \coherent achieves sensitivity independent of segment duration and surpasses the performance of the conventional incoherent-sum approach, through analytical arguments and numerical experiments.


\end{abstract}

\maketitle

\acrodef{ULBDM}{ultralight bosonic dark matter}
\acrodef{GW}{gravitational wave}
\acrodef{DM}{dark matter}
\acrodef{SNR}{signal-to-noise ratio}
\acrodef{PSD}{(two-sided) power spectral density}
\acrodef{CSD}{cross spectral density}

\acresetall

\section{Introduction} \label{sec:introduction}

The existence of \dm is firmly established by various observations, yet its origin and nature still remain unknown.
The allowed mass range of \dm particles spans many orders of magnitude: \dm could consist of ultralight bosons with masses as low as $10^{-22}$ eV or massive primordial black holes \cite{Bertone:2018krk}.
Many experiments have been dedicated to searching for \dm particles in the mass range from $10\,\mathrm{GeV}$ to $1\,\mathrm{TeV}$ \cite{Aprile:2018dbl,Ackermann:2015zua,Sirunyan:2017hci,Aaboud:2016tnv} as these particles would have been naturally produced with the observed abundance \cite{Bertone:2004pz}.
Despite the extensive efforts, no conclusive evidence has been found so far, motivating us to explore \dm particles in a broader parameter space.

Recently there have been various proposals to search for \ulbdm, whose mass ranges from $10^{-22}\,\mathrm{eV}$ to $1\,\mathrm{eV}$ \cite{Ferreira:2020fam}.
\ulbdm behaves as classical waves in the Galaxy due to its large occupation number.
They are nearly monochromatic waves whose spectra are concentrated around the Compton frequency $\fb \equiv \mb / (2 \pi)$ with a spectral width of the order of $\fb \vvir^2 \sim 10^{-6} \fb$, where $\mb$ is the mass of the boson and $\vvir$ is the virial velocity around the solar system \cite{Bertone:2004pz,Evans:2018bqy}. 
Throughout this paper, we apply the natural unit system, $\hbar=c=\varepsilon_0=1$.

For masses of $10^{-14}$--$10^{-11}\,\mathrm{eV}$, their frequencies fall within the sensitivity range of currently operating ground-based \gw detectors, such as LIGO \cite{LIGOScientific:2014pky}, Virgo \cite{VIRGO:2014yos}, and KAGRA \cite{KAGRA:2020tym}.
Hence, these detectors can be sensitive to \ulbdm as well as \gws.
For example, vector bosons such as dark photons can induce oscillatory forces on test masses in \gw detectors, potentially leading to nearly monochromatic signals in their outputs \cite{Pierce:2018xmy,Morisaki:2020gui}.
Searches based on this concept have already been conducted using data from LIGO, resulting in upper bounds on their coupling constants \cite{Guo:2019ker,LIGOScientific:2021ffg}.
A search using data from LISA Pathfinder \cite{McNamara:2008zz, Armano:2016bkm} has also been conducted \cite{Miller:2023kkd}.

A similar search has been performed using data from auxiliary channels of KAGRA \cite{KAGRA:2024ipf}.
The cryogenic mirrors and the other mirrors of KAGRA are made of sapphire and fused silica, respectively.
Consequently, they have different dark charges and experience varying accelerations from dark photons.
While these differences in acceleration are not detectable in the \gw channel, they can be measured in the auxiliary channels \cite{Michimura:2020vxn}.
The possibility to perform a search using auxiliary channels of LISA Pathfinder has also been discussed in \cite{Frerick:2023xnf}.

Scalar bosons, such as dilaton-like particles, and tensor bosons can also induce oscillatory forces on test masses, and searches for these particles can be conducted using \gw detectors \cite{Arvanitaki:2014faa,Morisaki:2018htj,Fukusumi:2023kqd,Armaleo:2020efr}.
Dilaton-like particles, in particular, can cause oscillatory changes in the fine structure constant and the electron rest mass, which can be investigated through various experiments and observations \cite{Grote:2019uvn,Vermeulen:2021epa,Gottel:2024cfj,Savalle:2020vgz,Hall:2022zvi,Stadnik:2014tta,Stadnik:2015kia,Stadnik:2015xbn,Aiello:2021wlp}.
Another type of scalar bosons, axions, can affect phase velocities of circularly polarized photons.
Such effects can be probed using existing \gw detectors with minor modifications \cite{Nagano:2019rbw,Nagano:2021kwx}.
Additionally, optical cavities specifically designed for axion searches are currently under development \cite{Obata:2018vvr,Oshima:2023csb,Heinze:2023nfb,Pandey:2024dcd}.

Due to the small but finite spectral width of a \dm field, the phase of signal slowly varies with time.
The variational time scale $\tauc$, called coherence time, is the inverse of the spectral width: $\tauc = 1 / (\fb \vvir^2)$.
Several data analysis methods have been proposed and employed to search for such a signal \cite{Guo:2019ker, Miller:2020vsl}.
In the previous studies, the authors employed an \incoherenth approach, where we divide data into segments with duration of $\tauc$, apply a Fourier transform to each segment, and incoherently sum up the signal power from all the segments \cite{Nakatsuka:2022gaf,KAGRA:2024ipf}.
If segment durations are longer than $\tauc$, the minimum detectable signal amplitude with this detection statistic, referred to as \incoherent, decreases with $\tauc$ and the total observation time $\Ttot$ in proportion to $\tauc^{-1/4} \Ttot^{-1/4}$ \cite{Morisaki:2018htj,Nakatsuka:2022gaf}.

On the other hand, detectors are not always operating stably, and obtaining contiguous segments longer than $\tauc$ can be impractical.
For example, the durations of contiguous data segments of KAGRA in the LIGO-Virgo-KAGRA's third observing run were at most 7 hours \cite{KAGRA:2022fgc} while $\tauc$ becomes around a day for $\fb=10\,\mathrm{Hz}$.
The coherence time becomes even longer at lower frequencies, where space-based detectors such as LISA~\cite{LISA:2017pwj} and DECIGO~\cite{Kawamura:2020pcg} are sensitive.
Furthermore, even if contiguous data of that duration are available, noise properties generally vary with time, and they may not be treated as stationary over a long segment.
Analyzing segments shorter than $\tauc$ effectively reduces the coherence time, degrading the sensitivity of \incoherent.

In this paper, it is pointed out that a \dm signal can be optimally extracted from data segments of arbitrary durations by exploiting expected signal correlations.
The proposed detection statistic, referred to as \coherent, is a quadratic form of Fourier components of segments, whose weights are computed with noise and signal correlations. 
It is shown that the minimum detectable signal amplitude with this method decreases in proportion to $\tauc^{-1/4} \Ttot^{-1/4}$ when $\Ttot > \tauc$ and does not depend on the segment lengths, resolving limitations caused by detector instabilities or non-stationary instrumental noise.
Our arguments generally apply to \ulbdm models in which the field interacts linearly with detectors, encompassing all the examples mentioned above.

This paper is organized as follows.
In Sec. \ref{sec:correlation}, formula for calculating signal correlations are derived and applied to specific \ulbdm models.
In Sec. \ref{sec:optimal}, a novel detection statistic is derived and the dependence of its sensitivity is studied both with analytical arguments and numerical simulations.
In reality, the mass of \dm particles is not known a priori, and a search needs to be conducted over a grid of \dm mass values.
In Sec. \ref{sec:template_bank}, a criterion for the grid spacing is derived to ensure that any potential signals are captured.
In Sec. \ref{sec:conclusion}, the results are summarized and future improvements of the method are discussed.

\section{Signal correlation} \label{sec:correlation}


Our optimal detection method requires a clear understanding of signal correlations across different data segments.
Therefore, our goal in this section is to calculate signal correlation between different data segments.  
For simplicity, we focus exclusively on \ulbdm models, where detectors are linearly correlated with \ulbdm fields and/or their spatial derivatives.
We assume that time-series data from a single or multiple detectors are available and data from each detector are divided into segments.
Each segment is specified by a pair of indexes $S = (I, s_I)$, where $I$ indexes detectors and $s_I$ indexes segments of the $I$--th detector.
Start time and duration of the segment $S$ is denoted by $t_S$ and $T_S$, respectively.

Each segment is multiplied by a window function $w(t - \bar{t}_S, S)$ and Fourier transformed.
Here, $w(t, S)$ is defined for any real value of $t$ and non-zero only in an interval with the length of $T_S$.
$w(t, S)$ is centered at $t=0$ so that
\begin{equation}
    \int^{\infty}_{-\infty} dt \, t \, w(t, S) = 0,
\end{equation}
and $\bar{t}_S$ is a time shift so that $w(t - \bar{t}_S, S)$ is non-zero only for $t_S \leq t \leq t_S + T_S$.
Each detector output contains signal time series, $h(t, I)$.
The Fourier transform of $h(t, I)$ for the segment $S$ is denoted by $\tilde{h}(f, S)$:
\begin{equation}
    \begin{aligned}
        &\tilde{h}(f, S) \equiv \int^{\infty}_{-\infty} dt w(t - \bar{t}_{S}, S) h(t, I) \e^{-2 \pi \iu f (t - t_{S})}.
    \end{aligned} \label{eq:hfS}
\end{equation}

The target quantity we aim to calculate is the signal correlation matrix $\scov$, whose component is given by
\begin{equation}
    \epsilon^2 \scov_{fS, f'S'} \equiv
    \left<\tilde{h}(f, S) \tilde{h}^\ast(f', S')\right>. \label{eq:cov_def}
\end{equation}
Here, the row or column of the matrix is specified by a pair of frequency $f$ and segment index $S$, and $\epsilon$ abstractly represents the coupling constant of \ulbdm.
The angle brackets denote an ensemble average over different signal realizations.
Since $\left<\tilde{h}(f, S)\right>=0$, the correlation matrix is the same as the covariance matrix.

\subsection{Spectral densities and correlation functions} \label{sec:psd}

First, we calculate the spectral densities and correlation functions of an \ulbdm field, which will be defined later.
An \ulbdm field at time $t$ and spatial position $\vec{x}$ within the Galaxy is modeled as a superposition of $N~(\gg 1)$ particle waves:
\begin{equation}
    \begin{aligned}
        &\bm{\mathcal{E}}(t, \vec{x}) = \\
        &~~\frac{A}{\sqrt{N}} \sum_{n=0}^{N-1}  \bm{e}_n \cos \left(2 \pi \fobs(v_n) t - 2 \pi \fb \vec{v}_n \cdot \vec{x} + \theta_n\right).
    \end{aligned}
\end{equation}
$\bm{\mathcal{E}}(t, \vec{x})$ can represent a scalar, vector, or tensor field depending on the \ulbdm model being considered.
$A$ represents the overall amplitude, and $\bm{e}_n$, $\vec{v}_n$ and $\theta_n$ represent the polarization, velocity and phase of the $n$-th particle respectively.
Each $\theta_n$ is statistically independent and follows a uniform distribution in the interval $[0, 2 \pi)$.
While $\vec{v}_n$ is often assumed to follow the standard halo model distribution, which is introduced later, we will consider a general distribution for now.
Each particle wave oscillates at the frequency of $\fb = \mb / (2 \pi)$ in its rest frame, and the frequency in the observer's frame where it has velocity $v$ is given by
\begin{equation}
    \fobs(v) \equiv \fb \left(1 + \frac{v^2}{2}\right).
\end{equation}

In the following calculations, we set $\vec{x} = \vec{0}$ and omit $\vec{x}$ from the input of $\bm{\mathcal{E}}(t, \vec{x})$.
This simplification is justified by the following considerations: First, the wavelength of an \ulbdm field is typically much larger than the separation between different detectors. 
The wavelength is roughly given by $1/(\fb \vvir) \sim 3 \times 10^6 (100\,\mathrm{Hz}/\fb) \,\mathrm{km}$.
For example, it is much larger than the separation between the two LIGO detectors, which is about $3 \times 10^3\,\mathrm{km}$, if $\fb$ falls within the frequency range of LIGO.
As a result, phase differences between signals at different detector locations are negligible, allowing us to treat all the detectors as being at the same position.

Second, velocities of detectors are so small that their positions can be considered constant over time.
We define our coordinate system to be a rest frame of the Sun, and velocities are measured relative to the Sun.
The velocity of a detector, $v_{\mathrm{det}}$, is typically much smaller than $\vvir$, and the distance a detector travels over the coherence time, $\tauc = 1 / (\fb \vvir^2)$, is much smaller than the wavelength of the \ulbdm field: $v_{\mathrm{det}} \tauc \ll 1 / (\fb \vvir)$.
Later in this section, we will show that the signal correlation between segments vanishes when their time separation is much greater than $\tauc$.
Since we do not need to accurately calculate vanishing components of the signal covariance matrix, we can approximate detectors' positions as fixed over time.
Since a phase factor arising from a constant detector position is canceled out in the following calculations, setting $\vec{x} = \vec{0}$ does not affect the results.

The Fourier transform of $\bm{\mathcal{E}}(t)$ for a positive frequency $f$ is given by
\begin{align}
    \tilde{\bm{\mathcal{E}}}(f) &\equiv \int^{\infty}_{-\infty} dt \bm{\mathcal{E}}(t) \e^{-2 \pi \iu f t} \\
    &= \frac{A}{2 \sqrt{N}} \sum_{n=0}^{N-1} \bm{e}_n \e^{\iu \theta_n} \delta \left(f - \fobs(v_n)\right),
\end{align}
and its value for a negative frequency $f$ can be obtained by $\tilde{\bm{\mathcal{E}}}(f) = \tilde{\bm{\mathcal{E}}}^\ast(-f)$.
Given each constant phase $\theta_n$ is statistically independent, the correlation of $\tilde{\bm{\mathcal{E}}}(f)$ at different frequencies, $f$ and $f'$, is given by
\begin{equation}
    \left<\tilde{\bm{\mathcal{E}}}(f) \otimes \tilde{\bm{\mathcal{E}}}^\ast(f')\right> = \bm{P}(f) \delta(f - f'), \label{eq:def_of_psd}
\end{equation}
where $\bm{P}(f)$ is the \psd of $\bm{\mathcal{E}}(f)$ and can be calculated as follows,
\begin{equation}
    \bm{P}(f) = \frac{A^2}{4 N} \sum_{n=0}^{N-1} \bm{e}_n \otimes \bm{e}_n \delta \left(|f| - \fobs(v_n)\right), \label{eq:psd_montecarlo}
\end{equation}
and $\bm{X} \otimes \bm{Y}$ denotes a tensor product of arbitrary tensors, $\bm{X}$ and $\bm{Y}$, whose components are given by the products of their components, $\left[\bm{X} \otimes \bm{Y}\right]_{ijk \dots abc \dots} = \left[\bm{X}\right]_{ijk \dots} \left[\bm{Y}\right]_{abc \dots}$.


In some \ulbdm models, detectors are coupled with spatial derivatives of $\bm{\mathcal{E}}(t, \vec{x})$. 
Let $\tilde{\bm{\mathcal{E}}}_{i_0 \cdots i_{L-1}}(f)$ denote the Fourier transform of the $L$-th spatial derivative,
\begin{equation}
    \begin{aligned}
        &\tilde{\bm{\mathcal{E}}}_{i_0 \cdots i_{L-1}}(f) \equiv \\
        &~~ \int^{\infty}_{-\infty} dt \partial_{i_0} \cdots \partial_{i_{L-1}} \bm{\mathcal{E}}(t, \vec{x})\bigg|_{\vec{x}=\vec{0}} \e^{-2 \pi \iu f t}.
    \end{aligned}
\end{equation}
Their correlation between different frequencies is given by
\begin{equation}
    \begin{aligned}
        &\left<\tilde{\bm{\mathcal{E}}}_{i_0 \cdots i_{L-1}}(f) \otimes \tilde{\bm{\mathcal{E}}}^\ast_{j_0 \cdots j_{M-1}}(f')\right> = \\
        &~~(-1)^L \bm{P}_{i_0 \cdots i_{L-1} j_0 \cdots j_{M-1}}(f) \delta(f - f'), \label{eq:def_of_psd_der}
    \end{aligned}
\end{equation}
where $\bm{P}_{i_0 \cdots i_{L-1}}(f)$ is the cross spectral density of $\tilde{\bm{\mathcal{E}}}(f)$ and $\tilde{\bm{\mathcal{E}}}_{i_0 \cdots i_{L-1}}(f)$ and can be computed as follows,
\begin{equation}
    \begin{aligned}
        &\bm{P}_{i_0 \cdots i_{L-1}}(f) = \frac{A^2}{4 N} \left(2 \pi \iu \fb \frac{f}{|f|}\right)^L \sum_{n=0}^{N-1} \bigg[ \\
        &~~\left[\vec{v}_n\right]_{i_0} \cdots \left[\vec{v}_n\right]_{i_{L-1}} \bm{e}_n \otimes \bm{e}_n \delta \left(|f| - \fobs(v_n)\right) \bigg]. \label{eq:psd_der_montecarlo}
    \end{aligned}
\end{equation}

$\bm{P}(f)$ and $\bm{P}_{i_0 \cdots i_{L-1}}(f)$ are dependent on velocity distribution of \ulbdm particles, denoted by $p(\vec{v})$, and the expectation value of the tensor product of polarizations, $\left< \bm{e} \otimes \bm{e} \right>_{\vec{v}}$.
The latter quantity can depend on velocity, and hence it has a subscript $\vec{v}$.
$\bm{P}(f)$ and $\bm{P}_{i_0 \cdots i_{L-1}}(f)$ can also be calculated with the following integrals,
\begin{align}
    &\begin{aligned}
        \bm{P}(f) = &\frac{A^2}{4} \Theta(\left|f\right| - \fb) \left(v(f)\right)^2 \left|\frac{dv}{df}\right| \times \\
        &\int d^2 \Omega_{\hat{\vec{v}}} p\left(v(f) \hat{\vec{v}}\right) \left< \bm{e} \otimes \bm{e} \right>_{v(f) \hat{\vec{v}}}, \\
    \end{aligned} \\
    &\begin{aligned}
        &\bm{P}_{i_0 \cdots i_{L-1}}(f) = \\
        &~~\frac{A^2}{4} \Theta(\left|f\right| - \fb) \left(2 \pi \iu \fb \frac{f}{|f|}\right)^L \left(v(f)\right)^{L + 2} \left|\frac{dv}{df}\right| \times \\
        &~~\int d^2 \Omega_{\hat{\vec{v}}}  p\left(v(f) \hat{\vec{v}}\right) \left[\hat{\vec{v}}\right]_{i_0} \cdots \left[\hat{\vec{v}}\right]_{i_{L-1}} \left< \bm{e} \otimes \bm{e} \right>_{v(f) \hat{\vec{v}}},
    \end{aligned}
\end{align}
where
\begin{equation}
    v(f) = \sqrt{2 \left(\frac{\left|f\right|}{\fb} - 1\right)}, \label{eq:def_of_v}
\end{equation}
$\hat{\vec{v}}$ is a unit vector, and $\Theta(x)$ is the Heaviside step function.

We consider a special case where $\left< \bm{e} \otimes \bm{e} \right>_{\vec{v}}$ is independent of velocity, $\left< \bm{e} \otimes \bm{e} \right>_{\vec{v}} \equiv \bm{E}$, and velocities follow the standard halo model distribution~\cite{Bertone:2004pz,Evans:2018bqy},
\begin{equation}
    \pSHM (\vec{v}) = \frac{1}{\left(\pi \vvir^2\right)^{3/2}} \e^{-\frac{(\vec{v} + \vec{V})^2}{\vvir^2}}.
\end{equation}
Here, $\vec{V}$ represents the velocity of the Sun relative to the Galactic center.
The assumption regarding polarizations holds in the case of scalar particles, where $\bm{E}=1$.
It also applies to vector particles with an isotropic distribution of polarizations, which was assumed, {\it e.g.}, in Refs.~\cite{Pierce:2018xmy,Guo:2019ker,LIGOScientific:2021ffg,KAGRA:2024ipf}.
In this case, $\bm{E}$ is a unit $3 \times 3$ matrix.
We refer to this case as the standard-halo-model (SHM) case.

In the SHM case, $\bm{P}(f)$ and $\bm{P}_{i_0 \cdots i_{L-1}}(f)$ can be calculated analytically.
Let $x(f)$ be defined as 
\begin{equation}
    x(f) \equiv \frac{2V}{\vvir^2} v(f),
\end{equation}
where $V=|\vec{V}|$, and $\delta^{\perp}_{ij}$ represents a projection operator orthogonal to $\vec{V}$,
\begin{equation}
    \delta^{\perp}_{ij} \equiv \delta_{ij} - \frac{V_i V_j}{V^2},
\end{equation}
where $\delta_{ij}$ represents the Kronecker delta.
When calculating $\bm{P}_{i_0 \cdots i_{L-1}}(f)$, it is helpful to observe that $\bm{P}_{i_0 \cdots i_{L-1}}(f)$ is a symmetric tensor constructed from $\delta^{\perp}_{ij}$ and $V_i$.
The analytical forms of $\bm{P}(f)$ and $\bm{P}_{i_0 \cdots i_{L-1}}(f)$ are given as follows:
\begin{widetext}
    \begin{align}
        &\bm{P}(f) = \frac{A^2 \vvir^3}{8 \sqrt{\pi} V^3} \Theta(\left|f\right| - \fb) \bm{E} x\left|\frac{dx}{df}\right| \e^{-\frac{v^2 + V^2}{\vvir^2}} \sinh x, \label{eq:analytic_p} \\
        &\bm{P}_i(f) = \iu \frac{\sqrt{\pi} A^2 \fb \vvir^5}{8 V^4} \frac{f}{|f|} \Theta(\left|f\right| - \fb) \bm{E} x\left|\frac{dx}{df}\right| \e^{-\frac{v^2 + V^2}{\vvir^2}} \left(-x \cosh x + \sinh x\right) \frac{V_i}{V}, \label{eq:analytic_pi} \\
        &\bm{P}_{ij}(f) = \frac{\pi^{\frac{3}{2}} A^2 \fb^2 \vvir^7}{8 V^5} \Theta(\left|f\right| - \fb) \bm{E} x \left|\frac{dx}{df}\right| \e^{-\frac{v^2 + V^2}{\vvir^2}} \bigg[ \nonumber \\
        &~~\left(-x \cosh x + \sinh x\right) \delta^{\perp}_{ij} + \left(2 x \cosh x - (2 + x^2) \sinh x\right) \frac{V_i V_j}{V^2} \bigg], \label{eq:analytic_pij} \\
        &\bm{P}_{ijk}(f) = \iu \frac{\pi^{\frac{5}{2}} A^2 \fb^3 \vvir^9}{8 V^6} \frac{f}{|f|} \Theta(\left|f\right| - \fb) \bm{E} x \left|\frac{dx}{df}\right| \e^{-\frac{v^2 + V^2}{\vvir^2}} \bigg[ \nonumber \\
        &~~\left(-3 x \cosh x + (3 + x^2) \sinh x\right) \left(\delta^\perp_{ij} \frac{V_k}{V} + \delta^\perp_{ik} \frac{V_j}{V} + \delta^\perp_{jk} \frac{V_i}{V}\right) + \left(x (6 + x^2) \cosh x - 3 (2 + x^2) \sinh x\right) \frac{V_i V_j V_k}{V^3} \bigg], \label{eq:analytic_pijk} \\
        &\bm{P}_{ijkl}(f) = \frac{\pi^{\frac{7}{2}} A^2 \fb^4 \vvir^{11}}{8 V^7} \Theta(\left|f\right| - \fb) \bm{E} x \left|\frac{dx}{df}\right| \e^{-\frac{v^2 + V^2}{\vvir^2}} \bigg[ \nonumber \\
        &~~\left(-3 x \cosh x + (3 + x^2) \sinh x\right) \left(\delta^{\perp}_{ij} \delta^{\perp}_{kl} + \delta^{\perp}_{ik} \delta^{\perp}_{jl} + \delta^{\perp}_{il} \delta^{\perp}_{jk}\right) + \nonumber \\
        &~~\left(x (12 + x^2) \cosh x - (12 + 5 x^2) \sinh x\right) \left(
        \delta^{\perp}_{ij} \frac{V_k V_l}{V^2} + \delta^{\perp}_{ik} \frac{V_j V_l}{V^2} + \delta^{\perp}_{il} \frac{V_j V_k}{V^2} + \delta^{\perp}_{jk} \frac{V_i V_l}{V^2} + \delta^{\perp}_{jl} \frac{V_i V_k}{V^2} + \delta^{\perp}_{kl} \frac{V_i V_j}{V^2}
        \right) + \nonumber \\
        &~~\left(-4 x (6 + x^2) \cosh x + (24 + 12 x^2 + x^4) \sinh x\right) \frac{V_i V_j V_k V_l}{V^4} \bigg]. \label{eq:analytic_pijkl}
    \end{align}
\end{widetext}
Here we consider spatial derivatives up to the fourth order since they are sufficient for the models we consider in this paper.

The auto-correlation function of $\bm{\mathcal{E}}(t)$ and cross-correlation function between $\bm{\mathcal{E}}(t)$ and $\bm{\mathcal{E}}_{i_0 \cdots i_{L-1}}(t)$ are obtained by taking the inverse Fourier transform of $\bm{P}(f)$ and $\bm{P}_{i_0 \cdots i_{L-1}}(f)$ respectively,
\begin{align}
    &\bm{R}(t) \equiv \int^\infty_{-\infty} df \bm{P}(f) \e^{2 \pi \iu f t}, \\
    &\bm{R}_{i_0 \cdots i_{L-1}}(t) \equiv \int^\infty_{-\infty} df \bm{P}_{i_0 \cdots i_{L-1}}(f) \e^{2 \pi \iu f t}.
\end{align}
In addition, we define the following complex functions,
\begin{align}
    &\bm{R}^{\mathrm{c}}(t) \equiv \int^\infty_{0} df \bm{P}(f) \e^{2 \pi \iu f t}, \label{eq:auto_corr_complex} \\
    &\bm{R}^{\mathrm{c}}_{i_0 \cdots i_{L-1}}(t) \equiv \int^\infty_{0} df \bm{P}_{i_0 \cdots i_{L-1}}(f) \e^{2 \pi \iu f t}. \label{eq:cross_corr_complex}
\end{align}
Since $\bm{P}(-f) = \bm{P}(f)$ and $\bm{P}_{i_0 \cdots i_{L-1}}(-f) = \bm{P}_{i_0 \cdots i_{L-1}}^\ast(f)$, the correlation functions are related to these complex functions as follows,
\begin{align}
    &\bm{R}(t) \equiv 2 \Re \left[\bm{R}^{\mathrm{c}}(t)\right], \\
    &\bm{R}_{i_0 \cdots i_{L-1}}(t) \equiv 2 \Re \left[\bm{R}^{\mathrm{c}}_{i_0 \cdots i_{L-1}}(t)\right].
\end{align}

In the SHM case, their analytical forms are given as follows:
\begin{widetext}
    \begin{align}
        &\bm{R}^{\mathrm{c}}(t) = \frac{A^2 \vvir^3}{4 V^3} \bm{E} \e^{-\frac{V^2}{\vvir^2} + 2 \pi \iu \fb t} y^{\frac{3}{2}} \e^y, \label{eq:analytic_r} \\
        &\bm{R}^{\mathrm{c}}_i(t) = -\iu \frac{\pi A^2 \fb \vvir^5}{2 V^4} \bm{E} \e^{-\frac{V^2}{\vvir^2} + 2 \pi \iu \fb t} y^{\frac{5}{2}} \e^y \frac{V_i}{V}, \label{eq:analytic_ri} \\
        &\bm{R}^{\mathrm{c}}_{ij}(t) = - \frac{\pi^2 A^2 \fb^2 \vvir^7}{2 V^5} \bm{E} e^{-\frac{V^2}{\vvir^2} + 2 \pi \iu \fb t} y^{\frac{5}{2}} \e^{y} \left[\delta^{\perp}_{ij} + (1 + 2 y) \frac{V_i V_j}{V^2} \right], \label{eq:analytic_rij} \\
        &\bm{R}^{\mathrm{c}}_{ijk}(t) = \iu \frac{\pi^{3} A^2 \fb^3 \vvir^9}{V^6} \bm{E} e^{-\frac{V^2}{\vvir^2} + 2 \pi \iu \fb t} y^{\frac{7}{2}} \e^y \left[\delta^\perp_{ij} \frac{V_k}{V} + \delta^\perp_{ik} \frac{V_j}{V} + \delta^\perp_{jk} \frac{V_i}{V} + (3 + 2 y) \frac{V_i V_j V_k}{V^3} \right], \label{eq:analytic_rijk} \\
        &\bm{R}^{\mathrm{c}}_{ijkl}(t) = \frac{\pi^{4} A^2 \fb^4 \vvir^{11}}{V^7} \bm{E} e^{-\frac{V^2}{\vvir^2} + 2 \pi \iu \fb t} y^{\frac{7}{2}} \e^y \bigg[ \left(\delta^{\perp}_{ij} \delta^{\perp}_{kl} + \delta^{\perp}_{ik} \delta^{\perp}_{jl} + \delta^{\perp}_{il} \delta^{\perp}_{jk}\right) + \nonumber \\
        &~~(1 + 2 y) \left(
        \delta^{\perp}_{ij} \frac{V_k V_l}{V^2} + \delta^{\perp}_{ik} \frac{V_j V_l}{V^2} + \delta^{\perp}_{il} \frac{V_j V_k}{V^2} + \delta^{\perp}_{jk} \frac{V_i V_l}{V^2} + \delta^{\perp}_{jl} \frac{V_i V_k}{V^2} + \delta^{\perp}_{kl} \frac{V_i V_j}{V^2}
        \right) + (3 + 12 y + 4 y^2) \frac{V_i V_j V_k V_l}{V^4} \bigg], \label{eq:analytic_rijkl}
    \end{align}
    where
    \begin{equation}
        y(t) \equiv \frac{V^2}{\vvir^2 (1 - \iu \pi \vvir^2 \fb t)}.
    \end{equation}
\end{widetext}
Each function is proportional to a positive power of $y(t)$, and $|y(t)|$ decays as $\propto \tauc / |t|$ for $|t| \gg \tauc$.
The timescale for the decay of the correlation functions is approximately $\sim \tauc$ for general \dm velocity distributions as long as the distribution is concentrated within $v \lesssim \vvir$.
Additionally, it is evident that correlation functions with more spatial derivatives involve higher powers of $y(t)$ and decay more rapidly for $|t| \gg \tauc$.
A spatial derivative corresponds to multiplying by $2 \pi \fb v(f) \hat{\vec{v}}$ in the frequency domain, which broadens the spectra and leads to faster decay of the correlation functions.

\subsection{Correlation between different segments}

For models of our interest, the signal time series $h(t)$ is a linear combination of the field $\mathcal{E}$ and/or its spatial derivatives, each of which is multiplied by a time-dependent detector response factor. To simplify the calculation of signal correlation, we assume that $T_{S}$ is much shorter than the variational time scales of detector responses, and detector responses are approximated as constants over each segment. 
For example, this assumption is valid for ground-based \gw detectors as long as segment lengths are much shorter than a day. Under this assumption, $\tilde{h}(f, S)$ is approximated as a linear combination of the following quantities,
\begin{align}
    &\tilde{\bm{\mathcal{E}}}(f, S) \equiv \int^{\infty}_{-\infty} dt w(t - \bar{t}_S, S) \bm{\mathcal{E}}(t) \e^{-2 \pi \iu f (t - t_{S})}, \\
    &\begin{aligned}
        &\tilde{\bm{\mathcal{E}}}_{i_0 \cdots i_{L-1}}(f, S) \equiv \int^{\infty}_{-\infty} dt \bigg[ \\
        &~ w(t - \bar{t}_S, S) \nabla_{i_0} \cdots \nabla_{i_{L-1}} \bm{\mathcal{E}}(t, \vec{x})\bigg|_{\vec{x}=\vec{0}} \e^{-2 \pi \iu f (t - t_{S})}\bigg].
    \end{aligned}
\end{align}

Let $\tilde{w}(f, S)$ denote the Fourier transform of $w(t, S)$,
\begin{equation}
    \tilde{w}(f, S) \equiv \int^{\infty}_{-\infty} dt w(t, S) \e^{-2 \pi \iu f t}.
\end{equation}
$\tilde{\bm{\mathcal{E}}}(f, S)$ and $\tilde{\bm{\mathcal{E}}}_{i_0 \cdots i_{L-1}}(f, S)$ are related to $\tilde{\bm{\mathcal{E}}}(f)$ and $\tilde{\bm{\mathcal{E}}}_{i_0 \cdots i_{L-1}}(f)$ as follows:
\begin{align}
    &\begin{aligned}
        &\tilde{\bm{\mathcal{E}}}(f, S) = \\
        &~~\e^{-2 \pi \iu f (\bar{t}_S - t_S)} \int^{\infty}_{-\infty} df' \tilde{w}(f - f', S) \tilde{\bm{\mathcal{E}}}(f') \e^{2 \pi \iu f' \bar{t}_{S}}, \label{eq:ef_window}
    \end{aligned} \\
    &\begin{aligned}
        &\tilde{\bm{\mathcal{E}}}_{i_0 \cdots i_{L-1}}(f, S) = \e^{-2 \pi \iu f (\bar{t}_S - t_S)} \times \\
        &~~\int^{\infty}_{-\infty} df' \tilde{w}(f - f', S) \tilde{\bm{\mathcal{E}}}_{i_0 \cdots i_{L-1}}(f') \e^{2 \pi \iu f' \bar{t}_{S}}. \label{eq:deref_window}
    \end{aligned}
\end{align}
Then, $\scov_{fS, f'S'}$ is approximated as a linear combination of the following quantities:
\begin{align}
    &\begin{aligned}
        &\bm{\mathcal{C}} (f, f', S, S') \equiv \\
        &~\e^{2 \pi \iu f (\bar{t}_S - t_S) - 2 \pi \iu f' (\bar{t}_{S'} - t_{S'})} \left<\tilde{\bm{\mathcal{E}}}(f, S) \otimes \tilde{\bm{\mathcal{E}}}^\ast(f', S')\right>,
    \end{aligned} \\
    &\bm{\mathcal{C}}_{i_0 \cdots i_{L-1}} (f, f', S, S') \equiv \e^{2 \pi \iu f (\bar{t}_S - t_S) - 2 \pi \iu f' (\bar{t}_{S'} - t_{S'})} \times \nonumber \\
    &~\left<\tilde{\bm{\mathcal{E}}}(f, S) \otimes \tilde{\bm{\mathcal{E}}}^\ast_{i_0 \cdots i_{L-1}}(f', S')\right>,
\end{align}
where the phase factor $\e^{2 \pi \iu f (\bar{t}_S - t_S) - 2 \pi \iu f' (\bar{t}_{S'} - t_{S'})}$ is introduced to cancel out trivial phase shifts.
They are related to $\bm{P}(f)$ and $\bm{P}_{i_0\cdots i_{L-1}}(f)$ as follows:
\begin{align}
    &\begin{aligned}
        &\bm{\mathcal{C}}(f, f', S, S') = \int^\infty_{-\infty} df'' \bigg[ \\
        &~\tilde{w}(f - f'', S) \tilde{w}^\ast(f' - f'', S') \bm{P}(f'') \e^{2 \pi \iu f'' (\bar{t}_{S} - \bar{t}_{S'})}\bigg], \label{eq:C_with_P}
    \end{aligned} \\
    &\begin{aligned}
        &\bm{\mathcal{C}}_{i_0 \cdots i_{L-1}} (f, f', S, S') = \int^\infty_{-\infty} df'' \tilde{w}(f - f'', S) \times \\
        &~\tilde{w}^\ast(f' - f'', S') \bm{P}_{i_0\cdots i_{L-1}}(f'') \e^{2 \pi \iu f'' (\bar{t}_{S} - \bar{t}_{S'})}. \label{eq:Cder_with_Pder}
    \end{aligned}
\end{align}

Substituting \eqref{eq:psd_montecarlo} and \eqref{eq:psd_der_montecarlo} into Eqs. \eqref{eq:C_with_P} and \eqref{eq:Cder_with_Pder}, we obtain formulae to compute $\bm{\mathcal{C}} (f, f', S, S')$ and $\bm{\mathcal{C}}_{i_0 \cdots i_{L-1}} (f, f', S, S')$.
If the spectral leakage between positive and negative frequencies is negligible, their formulae for $f, f' > 0$ are given as follows:
\begin{align}
    &\bm{\mathcal{C}}(f, f', S, S') = \frac{A^2}{4 N} \sum_{n=0}^{N-1} \bigg[ \nonumber \\
    &~~~~\tilde{w}\left(f - \fobs(v_n), S\right) \tilde{w}^\ast\left(f' - \fobs(v_n), S'\right) \times \nonumber \\
    &~~~~\bm{e}_n \otimes \bm{e}_n \e^{2 \pi \iu \fobs(v_n) (\bar{t}_{S} - \bar{t}_{S'})}\bigg], \label{eq:C_montecalro} \\
    &\bm{\mathcal{C}}_{i_0 \cdots i_{L-1}} (f, f', S, S') = \left(2 \pi \iu \fb\right)^L \frac{A^2}{4 N} \sum_{n=0}^{N-1} \bigg[ \nonumber \\
    &~~~~ \tilde{w}\left(f - \fobs(v_n), S\right) \tilde{w}^\ast\left(f' - \fobs(v_n), S'\right) \times \nonumber \\
    &~~~~ \left[\vec{v}_n\right]_{i_0} \cdots \left[\vec{v}_n\right]_{i_{L-1}} \bm{e}_n \otimes \bm{e}_n \e^{2 \pi \iu \fobs(v_n) (\bar{t}_{S} - \bar{t}_{S'})} \bigg]. \label{eq:Cder_montecalro}
\end{align}
They can be evaluated via a Monte Carlo method, where we generate random velocities and polarizations of \ulbdm particles and numerically calculate the sums over $n$.

In the SHM case, we can reduce them to one-dimensional integrals by substituting Eqs. \eqref{eq:analytic_p}--\eqref{eq:analytic_pijkl} into Eqs. \eqref{eq:C_with_P} and \eqref{eq:Cder_with_Pder}:
\begin{widetext}
\begin{align}
    &\bm{\mathcal{C}} (f, f', S, S') = \frac{A^2 \vvir^3}{8 \sqrt{\pi} V^3} \bm{E} \e^{- \frac{V^2}{\vvir^2} + 2 \pi \iu \fb (\bar{t}_S - \bar{t}_{S'})} C_1, \label{eq:C0} \\
    &\bm{\mathcal{C}}_i (f, f', S, S') = \iu \frac{\sqrt{\pi} A^2 \fb \vvir^5}{8 V^4} \bm{E} \e^{- \frac{V^2}{\vvir^2} + 2 \pi \iu \fb (\bar{t}_S - \bar{t}_{S'})} (C_1 - C_2) \frac{V_i}{V}, \label{eq:C1} \\
    &\bm{\mathcal{C}}_{ij} (f, f', S, S') = \frac{\pi^{\frac{3}{2}} A^2 \fb^2 \vvir^7}{8 V^5} \bm{E} \e^{- \frac{V^2}{\vvir^2} + 2 \pi \iu \fb (\bar{t}_S - \bar{t}_{S'})} \bigg[ (C_1 - C_2) \delta^{\perp}_{ij} + (2 C_2 - 2 C_1 - C_3) \frac{V_i V_j}{V^2} \bigg], \label{eq:C2} \\
    &\bm{\mathcal{C}}_{ijk} (f, f', S, S') = \iu \frac{\pi^{\frac{5}{2}} A^2 \fb^3 \vvir^9}{8 V^6} \bm{E} \e^{- \frac{V^2}{\vvir^2} + 2 \pi \iu \fb (\bar{t}_S - \bar{t}_{S'})} \bigg[ \nonumber \\
    &~~(C_3 - 3 C_2 + 3 C_1) \left(\delta^{\perp}_{ij} \frac{V_k}{V} + \delta^{\perp}_{ik} \frac{V_j}{V} + \delta^{\perp}_{jk} \frac{V_i}{V}\right) + \left(C_4 - 3 C_3 + 6 C_2 - 6 C_1\right) \frac{V_i V_j V_k}{V^3} \bigg], \label{eq:C3} \\
    &\bm{\mathcal{C}}_{ijkl} (f, f', S, S') = \frac{\pi^{\frac{7}{2}} A^2 \fb^4 \vvir^{11}}{8 V^7} \bm{E} \e^{- \frac{V^2}{\vvir^2} + 2 \pi \iu \fb (\bar{t}_S - \bar{t}_{S'})} \bigg[ \nonumber \\
    &~~(C_3 - 3 C_2 + 3 C_1) \left(\delta^{\perp}_{ij} \delta^{\perp}_{kl} + \delta^{\perp}_{ik} \delta^{\perp}_{jl} + \delta^{\perp}_{il} \delta^{\perp}_{jk}\right) +  \nonumber \\
    &~~(C_4 - 5 C_3 + 12 C_2 - 12 C_1) \left(\delta^{\perp}_{ij} \frac{V_k V_l}{V^2} + \delta^{\perp}_{ik} \frac{V_j V_l}{V^2} + \delta^{\perp}_{il} \frac{V_j V_k}{V^2} + \delta^{\perp}_{jk} \frac{V_i V_l}{V^2} + \delta^{\perp}_{jl} \frac{V_i V_k}{V^2} + \delta^{\perp}_{kl} \frac{V_i V_j}{V^2}\right) + \nonumber \\
    &~~(C_5 - 4 C_4 + 12 C_3 - 24 C_2 + 24 C_1) \frac{V_i V_j V_k V_l}{V^4} \bigg], \label{eq:C4}
\end{align}
where
\begin{align}
    C_1 &\equiv \int_0^{\infty} dx \tilde{w}\left(f - \fobs \left(\frac{\vvir^2}{2 V} x\right), S\right) \tilde{w}^\ast\left(f' - \fobs \left(\frac{\vvir^2}{2 V} x\right), S'\right)  \e^{- x^2 / \left(4 y\left(\bar{t}_S - \bar{t}_{S'}\right)\right)} x \sinh x, \label{eq:C1_integral} \\
    C_2 &\equiv \int_0^{\infty} dx \tilde{w}\left(f - \fobs \left(\frac{\vvir^2}{2 V} x\right), S\right) \tilde{w}^\ast\left(f' - \fobs \left(\frac{\vvir^2}{2 V} x\right), S'\right)  \e^{- x^2 / \left(4 y\left(\bar{t}_S - \bar{t}_{S'}\right)\right)} x^2 \cosh x, \\
    C_3 &\equiv \int_0^{\infty} dx \tilde{w}\left(f - \fobs \left(\frac{\vvir^2}{2 V} x\right), S\right) \tilde{w}^\ast\left(f' - \fobs \left(\frac{\vvir^2}{2 V} x\right), S'\right)  \e^{- x^2 / \left(4 y\left(\bar{t}_S - \bar{t}_{S'}\right)\right)} x^3 \sinh x, \\
    C_4 &\equiv \int_0^{\infty} dx \tilde{w}\left(f - \fobs \left(\frac{\vvir^2}{2 V} x\right), S\right) \tilde{w}^\ast\left(f' - \fobs \left(\frac{\vvir^2}{2 V} x\right), S'\right)  \e^{- x^2 / \left(4 y\left(\bar{t}_S - \bar{t}_{S'}\right)\right)} x^4 \cosh x, \\
    C_5 &\equiv \int_0^{\infty} dx \tilde{w}\left(f - \fobs \left(\frac{\vvir^2}{2 V} x\right), S\right) \tilde{w}^\ast\left(f' - \fobs \left(\frac{\vvir^2}{2 V} x\right), S'\right)  \e^{- x^2 / \left(4 y\left(\bar{t}_S - \bar{t}_{S'}\right)\right)} x^5 \sinh x. \label{eq:C5_integral}
\end{align}

\end{widetext}

If $T_S,T_{S'} \ll \tauc$, the window functions are almost constant in the frequency span where $\bm{P}(f)$ or $\bm{P}_{i_0\cdots i_{L-1}}(f)$ is not negligible.
In this limit, $\bm{\mathcal{C}}(f, f', S, S')$ and $\bm{\mathcal{C}}_{i_0 \cdots i_{L-1}} (f, f', S, S')$ for $f, f' > 0$ are related to the complex correlation functions defined by Eqs. \eqref{eq:auto_corr_complex} and \eqref{eq:cross_corr_complex} as follows,
\begin{align}
    &\begin{aligned}
        &\bm{\mathcal{C}}(f, f', S, S') \simeq \\
        &~~\tilde{w}(f - \fb, S) \tilde{w}^\ast(f' - \fb, S') \bm{R}^{\mathrm{c}}(\bar{t}_S - \bar{t}_{S'}),
    \end{aligned} \label{eq:r_to_c} \\
    &\begin{aligned}
        &\bm{\mathcal{C}}_{i_0 \cdots i_{L-1}} (f, f', S, S') \simeq \\
        &~~\tilde{w}(f - \fb, S) \tilde{w}^\ast(f' - \fb, S') \bm{R}^{\mathrm{c}}_{i_0 \cdots i_{L-1}}(\bar{t}_S - \bar{t}_{S'}). \label{eq:rder_to_cder}
    \end{aligned}
\end{align}
Therefore, those quantities in the SHM case with $T_S,T_{S'} \ll \tauc$ can be analytically calculated with the above relations and Eqs. \eqref{eq:analytic_r}--\eqref{eq:analytic_rijkl}.

In summary, $\bm{\mathcal{C}}(f, f', S, S')$ and $\bm{\mathcal{C}}_{ij\cdots}(f, f', S, S')$ are calculated with one of the following methods: (1) For a general distribution of velocities and polarizations, they can be computed via a Monte Carlo method based on Eqs. \eqref{eq:C_montecalro} and \eqref{eq:Cder_montecalro}. (2) In the SHM case, they can be calculated using Eqs. \eqref{eq:C0}–\eqref{eq:C4}, with $C_1$--$C_5$ numerically computed through Eqs. \eqref{eq:C1_integral}–\eqref{eq:C5_integral}. (3) In the SHM case with $T_S,T_{S'} \ll \tauc$, they can be calculated analytically using Eqs. \eqref{eq:analytic_r}--\eqref{eq:analytic_rijkl}, \eqref{eq:r_to_c}, and \eqref{eq:rder_to_cder}.

\subsection{Specific models} \label{sec:specific}

Once $\bm{\mathcal{C}}(f, f', S, S')$ and $\bm{\mathcal{C}}_{ij\cdots}(f, f', S, S')$ are computed, the signal correlation matrix $\scov_{fS, f'S'}$ is obtained as their linear combination, whose coefficients depend on the detector responses.
Here we consider specific \ulbdm models and present the formulas to calculate $\scov_{fS, f'S'}$ using $\bm{\mathcal{C}}(f, f', S, S')$ and $\bm{\mathcal{C}}_{ij\cdots}(f, f', S, S')$.
In particular, we examine four \ulbdm models: axion, dilaton-like \dm, dark photon, and spin-2 \dm.
For scalar \dm, $\bm{\mathcal{C}}(f, f', S, S')$ and $\bm{\mathcal{C}}_{ij\cdots}(f, f', S, S')$ are scalar quantities, which we denote without boldface as $\mathcal{C}(f, f', S, S')$ and $\mathcal{C}_{ij\cdots}(f, f', S, S')$.
For vector or spin-2 \dm, these covariance functions are tensor quantities, with their components written as $\mathcal{C}_{|kl\cdots}(f, f', S, S') \equiv \left[\bm{\mathcal{C}}(f, f', S, S')\right]_{kl\cdots}$ and $\mathcal{C}_{ij\cdots|kl\cdots}(f, f', S, S') \equiv \left[\bm{\mathcal{C}}_{ij\cdots}(f, f', S, S')\right]_{kl\cdots}$.
For models involving time-varying detector responses, we introduce a reference time parameter, $t_S^{\mathrm{det}}$, at which the detector responses are evaluated for the segment $S$.

\subsubsection{Axion}

Axion is a pseudo-scalar particle that couples to photons through the Chern-Simons interaction
\begin{equation}
    \mathcal{L} \supset \frac{g_{a\gamma}}{4}a F_{\mu\nu}\tilde{F}^{\mu\nu},
\end{equation}
where $a(t, \vec{x})$ is the axion field, $F_{\mu\nu}$ is the electromagnetic tensor, $\tilde{F}^{\mu\nu} \equiv \epsilon^{\mu\nu\rho\sigma}F_{\rho\sigma}/2$ is its dual defined with the Levi-Civita antisymmetric tensor $\epsilon^{\mu\nu\rho\sigma}$ and $g_{a \gamma}$ is the axion-photon coupling constant.
In the background of an axion field, phase velocities of left-handed and right-handed photons with wave number $k$ are modulated as $c_{\mathrm{L}/\mathrm{R}} = 1 \mp \delta c(t)$, where~\cite{Carroll:1989vb,Carroll:1998zi}
\begin{equation}
    \delta c = \frac{g_{a \gamma} \partial_t a}{2 k}.
\end{equation}

Suppose that we have time-series data containing $\delta c$, that is, $h(t,I)=\delta c$, obtained from polarization measurements of laser light in a \gw detector \cite{Nagano:2019rbw,Nagano:2021kwx} or optical cavities designed for axion search \cite{Obata:2018vvr}, $\scov_{fS, f'S'}$ is given by
\begin{equation}
    \begin{aligned}
        &\scov_{fS, f'S'} = \\
        &~\e^{- 2 \pi \iu f (\bar{t}_S - t_S) + 2 \pi \iu f' (\bar{t}_{S'} - t_{S'})} \left(\frac{\pi \fb}{k}\right)^2 \mathcal{C}(f, f', S, S'), \label{eq:axion}
    \end{aligned}
\end{equation}
where we set $\epsilon = g_{a \gamma}$.

\subsubsection{Dilaton-like \dm}
Dilaton-like \dm is a scalar particle coupling to the Standard Model particles.
It is denoted by $\phi(t, \vec{x})$ and its interaction terms are given by~\cite{Damour:2010rp}
\begin{equation}
\begin{aligned}
    \mathcal{L} \supset \kappa \phi 
    \bigg[ &\frac{d_e}{4e^2}F_{\mu\nu}F^{\mu\nu} - \frac{d_g\beta_3}{2 g_3}G^a_{\mu\nu}G^{a\mu\nu} \\
    &-\sum_{i = e,u,d} (d_{m_i} + \gamma_{m_i} d_{g})
    m_i\bar{\psi}_i\psi_i
    \bigg],
\end{aligned}
\end{equation}
where $\kappa \equiv \sqrt{4\pi}/M_{\rm pl}$ with $M_{\rm pl}$ representing the Planck mass, $e$ is the electron charge, $g_3$ is the SU(3) gauge coupling, $\beta_3$ is the corresponding beta function, $G^a_{\mu\nu}$ is the gluon field strength, $\psi_i$ represents the fields for electrons or up/down quarks, $m_i$ represents their masses, and $\gamma_{m_i}$ are their anomalous dimensions.
At low energy, this scalar field changes the fine structure constant $\alpha$ and fermion masses as
\begin{align}
\alpha(\phi) &= (1 + d_e\kappa\phi)\alpha,\label{eq:alpha}\\
m_e(\phi) &= (1 + d_{m_e}\kappa\phi)m_e,\label{eq:me}\\
\left[m_{u,d}(\Lambda_3)\right](\phi) &= (1 + d_{m_{u,d}}\kappa\phi)m_{u,d}(\Lambda_3),
\end{align}
where $\Lambda_3$ is the QCD scale.

In the literature, the interaction between this scalar field and a \gw interferometer was considered and classified into two different contributions.
Refs.~\cite{Grote:2019uvn,Vermeulen:2021epa,Gottel:2024cfj} focused on oscillatory changes in $\alpha(\phi)$ and $m_e(\phi)$ that affect the refractive indexes and the thicknesses of the mirrors.
For a Michelson interferometer such as the GEO600 detector, the signal is given as~\cite{Grote:2019uvn,Vermeulen:2021epa}
\begin{equation}
    h(t,I) = \frac{\sqrt{2}}{L_I} \left(n_I - \frac{1}{2}\right) l_I (d_e + d_{m_e})\kappa\phi,
\end{equation}
where $n_I$ and $l_I$ are the refractive index and the thickness of the beamsplitter respectively, and $L_I$ represents the arm length.
By setting $\epsilon = d_e + d_{m_e}$, the signal covariance is then given as
\begin{equation}
    \begin{aligned}
        \scov_{fS, f'S'} = &\e^{- 2 \pi \iu f (\bar{t}_S - t_S) + 2 \pi \iu f' (\bar{t}_{S'} - t_{S'})} \times \\
        &\frac{2}{L_I L_{I'}} \left(n_I - \frac{1}{2}\right) l_I \left(n_{I'} - \frac{1}{2}\right) l_{I'}
        \kappa^2
        \mathcal{C}, \label{eq:cardiff}
    \end{aligned}
\end{equation}
where, as in the following, we omit the argument of the last term $(f, f', S, S')$.

On the other hand, Refs.~\cite{Morisaki:2018htj,Fukusumi:2023kqd} focused on signal caused by displacements of mirrors in a \gw detector.
Let $\hat{n}^i (t, I)$ and $\hat{m}^i (t, I)$ denote unit vectors along the arms of a \gw detector $I$. 
We then define $d^i(t, I)$ and $D^{ij}(t, I)$ as follows,
\begin{align}
    &d^i(t, I) \equiv \hat{n}^i (t, I) - \hat{m}^i(t, I), \\
    &D^{ij}(t, I) \equiv \hat{n}^i(t, I) \hat{n}^j(t, I) - \hat{m}^i(t, I) \hat{m}^j(t, I).
\end{align}
The signal is given by \cite{Morisaki:2018htj}
\begin{equation}
    \begin{aligned}
        &h(t, I) = \frac{d^\ast_g \kappa}{2} \bigg[ \\
        &~\frac{\sin^2 (\pi \fb L_I)}{\pi^2 \fb^2 L_I} d^i(t, I) \partial_i \phi + \frac{1}{2 \pi^2 \fb^2} D^{ij}(t, I) \partial_i \partial_j \phi \bigg],
    \end{aligned} \label{eq:dilaton_gw_signal}
\end{equation}
where $d^\ast_g$ denotes the coupling constant that is a linear combination of $d_g$, $d_e$ and $d_{m_i}$.
Note that $d_g$ dominantly contributes to $d_g^*$ as the mass of nucleons is mainly determined by the QCD interaction~\cite{Damour:2010rp}.
The first term in Eq. \eqref{eq:dilaton_gw_signal} arises because mirrors oscillate while laser light is traveling between the mirrors and it experiences apparent length changes even if the mirrors oscillate in a completely common way \cite{Morisaki:2018htj,Morisaki:2020gui}. 
We refer to this contribution as the effects of finite light-traveling time.
The second term arises because the forces acting on mirrors at different positions are slightly different and there are actual length changes between them.

Setting $\epsilon=d^\ast_g$, we have
\begin{align}
    &\scov_{fS, f'S'} = \frac{\kappa^2}{16 \pi^4 \fb^4} \e^{- 2 \pi \iu f (\bar{t}_S - t_S) + 2 \pi \iu f' (\bar{t}_{S'} - t_{S'})} \bigg[ \nonumber \\
    &~-\frac{4 \sin^2\left(\pi \fb L_{I}\right) \sin^2\left(\pi \fb L_{I'}\right)}{L_{I} L_{I'}} d^i d'^j \mathcal{C}_{ij} - \nonumber \\
    &~2 \left(\frac{\sin^2\left(\pi \fb L_{I}\right)}{L_{I}} d^i D'^{jk} - \frac{\sin^2\left(\pi \fb L_{I'}\right)}{L_{I'}} d'^i D^{jk}\right) \mathcal{C}_{ijk} + \nonumber \\
    &~D^{ij} D'^{kl} \mathcal{C}_{ijkl}\bigg], \label{eq:dilaton_gw}
\end{align}
where $d^i$, $d'^i$, $D^{ij}$, and $D'^{ij}$ represent $d^i(t_S^{\mathrm{det}}, I)$, $d^i(t_{S'}^{\mathrm{det}}, I')$, $D^{ij}(t_S^{\mathrm{det}}, I)$, and $D^{ij}(t_{S'}^{\mathrm{det}}, I')$ respectively.

\subsubsection{Dark photon}

Dark photon is a vector particle, denoted by $\vec{A}(t, \vec{x})$, which couples to the current of baryon number $B$ or baryon number minus lepton number $B - L$.
The signal in a \gw channel, caused by the mirror displacements induced by dark photon, is given by \cite{Morisaki:2020gui},
\begin{equation}
    \begin{aligned}
        &h(t, I) = \epsilon_D e \frac{Q_D}{M} \bigg[ \\
        &~~\frac{2 \sin^2\left(\pi \fb L_I\right)}{L_I} d^i(t, I) B_i + D^{ij}(t, I) \partial_i B_j \bigg], \label{eq:dark_photon_gw_signal}
    \end{aligned}
\end{equation}
where $D=B$ or $D=B-L$ depending on which model we consider, $\epsilon_D$ represents the coupling constant of dark photon normalized by the electron charge $e$,
$M$ and $Q_D$ represent the mass and dark charge of the mirrors respectively, and
\begin{equation}
    \vec{B}(t, \vec{x}) \equiv \int^t dt' \vec{A}(t', \vec{x}). 
\end{equation}
The first and second terms in Eq. \eqref{eq:dark_photon_gw_signal} represent the effects of finite light-traveling time and a signal caused by actual length changes respectively.
Given that the Fourier transform of $B_i(t, \vec{x})$ can be approximated as $\tilde{A}_i(f) / (2 \pi \iu \fb)$, $\scov_{fS, f'S'}$ is given by
\begin{align}
    &\scov_{fS, f'S'} = \frac{e^2}{4 \pi^2 \fb^2} \left(\frac{Q_D}{M}\right)^2 \e^{- 2 \pi \iu f (\bar{t}_S - t_S) + 2 \pi \iu f' (\bar{t}_{S'} - t_{S'})} \bigg[ \nonumber \\
    &~\frac{4 \sin^2\left(\pi \fb L_{I}\right) \sin^2\left(\pi \fb L_{I'}\right)}{L_{I} L_{I'}} d^i d'^j \mathcal{C}_{|ij} + \nonumber \\
    &~2 \left(\frac{\sin^2\left(\pi \fb L_{I}\right)}{L_{I}} d^i D'^{jk} - \frac{\sin^2\left(\pi \fb L_{I'}\right)}{L_{I'}} d'^i D^{jk}\right) \mathcal{C}_{j|ik} - \nonumber \\
    &~D^{ij} D'^{kl} \mathcal{C}_{ik|jl}\bigg], \label{eq:dark_photon_gw}
\end{align}
where we set $\epsilon=\epsilon_D$.

Signals are induced also in KAGRA's auxiliary channels, such as its differential Michelson arm length (MICH), power recycling cavity length (PRCL), and signal recycling cavity length (SRCL) channels.
Let $I$ label these auxiliary channels, and $\hat{\vec{n}}_{\mathrm{K}}$ and $\hat{\vec{m}}_{\mathrm{K}}$ denote unit vectors along the arms of KAGRA. 
We then define $d^i_{\mathrm{aux}}(t, I)$ as follows,
\begin{align}
    &d^i_{\mathrm{aux}}(t, I) \equiv \nonumber \\
    &~~\begin{cases}
        \hat{n}^i_{\mathrm{K}}(t) - \hat{m}^i_{\mathrm{K}}(t), & (I=\text{MICH}), \\
        \frac{1}{2} \left(\hat{n}^i_{\mathrm{K}}(t) + \hat{m}^i_{\mathrm{K}}(t)\right), & (I \in \{\text{PRCL, SRCL}\}).
    \end{cases}
\end{align}
The signal in the channel $I$ is then given by \cite{Michimura:2020vxn},
\begin{equation}
    h(t, I) = - \epsilon_D e \Delta \left(\frac{Q_D}{M}\right) d^i_{\mathrm{aux}}(t, I) B_i(t, \vec{x}),
\end{equation}
where $\Delta (Q_D/M)$ is the difference of $Q_D/M$ between sapphire and fused silica.
$\scov_{fS, f'S'}$ is given by
\begin{equation}
    \begin{aligned}
        \scov_{fS, f'S'} = &\e^{- 2 \pi \iu f (\bar{t}_S - t_S) + 2 \pi \iu f' (\bar{t}_{S'} - t_{S'})} \times \\
        &\left(\frac{e}{2 \pi \fb} \Delta \left(\frac{Q_D}{M}\right)\right)^2 d^i_{\mathrm{aux}} \mathcal{C}_{|ij} d'^j_{\mathrm{aux}},
    \end{aligned}
\end{equation}
where $d^i_{\mathrm{aux}}=d^i_{\mathrm{aux}}(t_S^{\mathrm{det}}, I)$ and $d'^i_{\mathrm{aux}}=d^i_{\mathrm{aux}}(t_{S'}^{\mathrm{det}}, I')$.

\subsubsection{Spin-2 \dm}

Spin-2 \dm, denoted by $\Phi_{ij}$, is a tensor field that universally couples to the matter fields as \cite{Aoki:2016zgp, Manita:2023mnc, Armaleo:2020efr},
\begin{equation}
\mathcal{L} \supset - \frac{\alpha}{2 M_{\rm Pl}}\Phi_{ij}T^{ij},
\end{equation}
where $\alpha$ is a dimensionless constant.
The signal in a \gw channel can be calculated using the same methods applied to \gws (for example, see Sec. 9.1 of \cite{Maggiore:2007ulw}), and it is given by
\begin{equation}
    h(t,D) = \frac{\alpha}{2 M_{\rm Pl}} \frac{\sin(2 \pi \fb L)}{2 \pi \fb L} D^{ij}(t, I)\Phi_{ij}.
\end{equation}
The response to \gws depends on their propagation direction when their frequency is comparable to or exceeds the inverse of the light travel time between the interferometer arms \cite{Rakhmanov:2008is}.
In contrast, the response to spin-2 \dm is direction-independent because of its non-relativistic particle velocities.
Setting $\epsilon = \alpha$, the signal correlation can be expressed as
\begin{equation}
\begin{aligned}
\scov_{fS, f'S'} = &\e^{- 2 \pi \iu f (\bar{t}_S - t_S) + 2 \pi \iu f' (\bar{t}_{S'} - t_{S'})} \times \\
&\frac{1}{4 M_{\rm Pl}^2} \left(\frac{\sin(2 \pi \fb L)}{2 \pi \fb L}\right)^2
D^{ij}\mathcal{C}_{|ijkl} D^{\prime kl}.\label{eq:cov_spin2}
\end{aligned}
\end{equation}

\begin{figure}[t]
	\centering
	\includegraphics[width=0.9\linewidth]{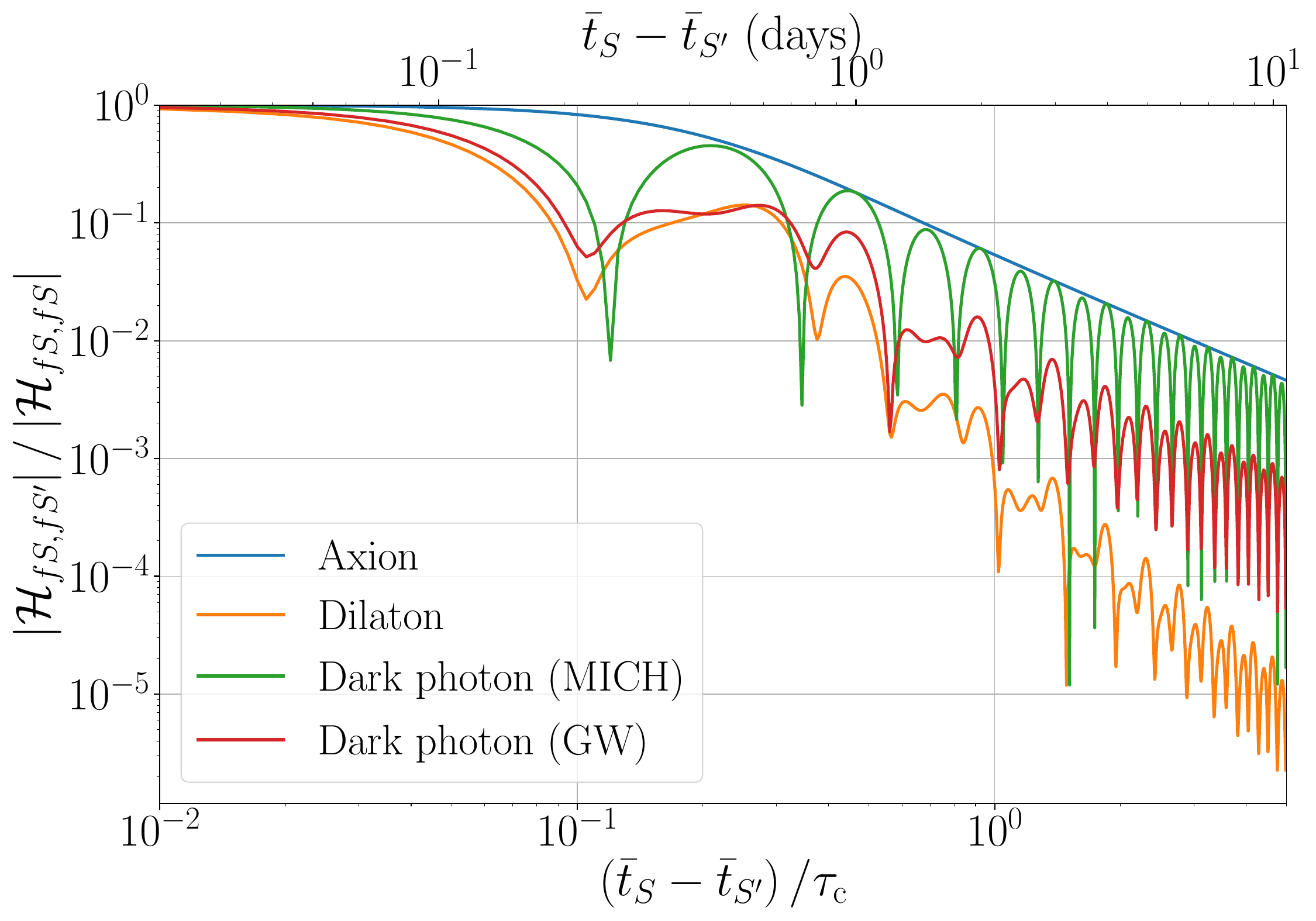}
    \caption{Normalized signal correlation, $\left|\scov_{fS, fS'} / \scov_{fS, fS}\right|$, as a function of $\bar{t}_{S} - \bar{t}_{S'}$, for the axion signal (blue), dilaton signal in the KAGRA's \gw channel (orange), and dark-photon signals in the KAGRA's MICH channel (green) and \gw channel (red). It is calculated with $f=\fb=10\,\mathrm{Hz}$, and for the SHM case in the limit of small segment durations, where Eqs. \eqref{eq:analytic_r}--\eqref{eq:analytic_rijkl}, \eqref{eq:r_to_c}, and \eqref{eq:rder_to_cder} are applicable.} \label{fig:correlation}
\end{figure}

Figure \ref{fig:correlation} shows the normalized signal correlation, $\left|\scov_{fS, fS'} / \scov_{fS, fS}\right|$, as a function of $\bar{t}_{S} - \bar{t}_{S'}$, for selected \dm models.
It is calculated with $f=\fb=10\,\mathrm{Hz}$, which corresponds to $\tau \simeq 10^{5}\,\mathrm{s} \simeq 1.2\,\mathrm{days}$.
It is computed for the SHM case, in the limit of small segment durations, where Eqs. \eqref{eq:analytic_r}--\eqref{eq:analytic_rijkl}, \eqref{eq:r_to_c}, and \eqref{eq:rder_to_cder} are applicable.
For dilaton-like \dm, we consider the signal caused by the displacements of the mirrors.
The normalized correlation for Eq.~\eqref{eq:cardiff} is the same as that of the axion signal.

As seen in the figure, the correlation starts to vanish when $\bar{t}_{S} - \bar{t}_{S'}$ gets comparable to $\tauc$.
In addition, the correlation decays more quickly for the dilaton and dark-photon signals of the \gw channel.
At this frequency, the terms proportional to $\mathcal{C}_{ijkl}$ and $\mathcal{C}_{ik|jl}$ are dominant in Eqs. \eqref{eq:dilaton_gw} and \eqref{eq:dark_photon_gw} respectively at $\bar{t}_{S} - \bar{t}_{S'} = 0$, which are proportional to $\bm{R}^{\mathrm{c}}_{ijkl}(\bar{t}_S - \bar{t}_{S'})$ and $\bm{R}^{\mathrm{c}}_{ik}(\bar{t}_S - \bar{t}_{S'})$ respectively.
On the other hand, the correlation for the axion or dark-photon signal of the MICH channel is proportional to $\bm{R}^{\mathrm{c}}(\bar{t}_S - \bar{t}_{S'})$, which decay more slowly than $\bm{R}^{\mathrm{c}}_{ijkl}(\bar{t}_S - \bar{t}_{S'})$ and $\bm{R}^{\mathrm{c}}_{ik}(\bar{t}_S - \bar{t}_{S'})$, as explained below Eqs. \eqref{eq:analytic_r}--\eqref{eq:analytic_rijkl}.
For models except for the axion model, the correlation function exhibits oscillations with the time scale of a day due to the time-varying detector responses.


\section{Optimal detection statistic}  \label{sec:optimal}

In this section, we derive an optimal detection statistic for an \ulbdm signal, taking into account signal correlations, and study its sensitivity.
Let $\vec{d}$ be a vector of $N$ Fourier data samples across different frequency bins, data segments, and detectors.
We model $\vec{d}$ as the sum of noise $\vec{n}$ and signal $\vec{h}$, where $\vec{n}$ is assumed to follow a circularly-symmetric Gaussian distribution with zero mean.
The noise's covariance matrix is denoted by $\ncov$.
If the noise time series can be modeled as a stationary random process with negligible correlations across different segments or detectors, $\ncov$ is diagonal, $\ncov_{fS, f'S'} = T_S P_{\mathrm{n}}(f, S) \delta_{ff'} \delta_{SS'}$, 
where $P_{\mathrm{n}}(f, S)$ represents the noise \psd and $\delta_{ff'}$ ($\delta_{SS'}$) is unity when $f = f'$ ($S = S'$) and zero otherwise.

An \ulbdm signal is a sum of contributions from many particle waves and, due to the central limit theorem, $\vec{h}$ approximately follows a circularly-symmetric Gaussian distribution with zero mean and the covariance matrix of $\epsilon^2 \scov$.
Under the assumption that noise and signal are statistically independent, $\vec{d}$ follows a circularly-symmetric Gaussian distribution with zero mean and the covariance matrix of $\dcov = \ncov + \epsilon^2 \scov$:
\begin{equation}
    p(\vec{d}) = \frac{1}{\pi^N \det \dcov} \e^{-\vec{d}^\dagger \dcov^{-1} \vec{d}}.
\end{equation}

\subsection{Derivation} \label{sec:optimal_derivation}

We assume that an optimal detection statistic, denoted by $\ds$, has a quadratic form of $\vec{d}$,
\begin{equation}
    \ds = \vec{d}^\dagger \mathcal{K} \vec{d} + \ds_0,
\end{equation}
where $\mathcal{K}$ is an Hermitian matrix and $\ds_0$ is a real constant.
This is a general quadratic form that is real and does not depend on the overall phase of $\vec{d}$.
The detection statistic employed in our previous studies is given by \cite{Nakatsuka:2022gaf,KAGRA:2024ipf},
\begin{equation}
    \dsinc = 2 \vec{d}^\dagger \ncov^{-1} \vec{d}, 
\end{equation}
which corresponds to $\mathcal{K}=2 \ncov^{-1}$ and $\ds_0=0$. 
This statistic, referred to as \incoherent, will be compared with the new detection statistic derived in this section.

We determine $\mathcal{K}$ by maximizing the following \snr,
\begin{equation}
    \mathrm{SNR} \equiv \frac{\left<\ds\right> - \left<\ds\right>|_{\epsilon = 0}}{\sqrt{\mathrm{Var}[\ds]|_{\epsilon=0}}}, \label{eq:SNR}
\end{equation}
where $\left<\cdot\right>$ and $\mathrm{Var}[\cdot]$ represent the expectation value and the variance respectively, and quantities subscripted by $\epsilon=0$ are calculated under the assumption that data do not contain any signals.
$\left<\ds\right>$ and $\mathrm{Var}[\ds]$ are given by
\begin{equation}
    \left<\ds\right> = \mathrm{Tr}\left[\mathcal{K} \dcov\right] + \rho_0,~~~\mathrm{Var}[\ds] = \mathrm{Tr}\left[\mathcal{K} \dcov \mathcal{K} \dcov\right].
\end{equation}
Substituting them into Eq. \eqref{eq:SNR}, we obtain
\begin{align}
    \mathrm{SNR} &= \epsilon^2 \frac{\mathrm{Tr}\left[\mathcal{K}\scov\right]}{\sqrt{\mathrm{Tr}\left[\mathcal{K} \ncov \mathcal{K} \ncov\right]}} \\
    &= \epsilon^2 \frac{\mathrm{Tr}\left[\mathcal{K}' \ncov^{-\frac{1}{2}} \scov \ncov^{-\frac{1}{2}}\right]}{\sqrt{\mathrm{Tr}\left[\left(\mathcal{K}'\right)^2\right]}},
\end{align}
where $\mathcal{K}' \equiv \ncov^{\frac{1}{2}} \mathcal{K} \ncov^{\frac{1}{2}}$.

Applying the Cauchy-Schwarz inequality, we can prove that the \snr is bounded by
\begin{equation}
    \mathrm{SNR} \leq \epsilon^2 \sqrt{\mathrm{Tr}\left[\left(\ncov^{-1} \scov\right)^2\right]},
\end{equation}
and the equality holds when $\mathcal{K}' \propto \ncov^{-\frac{1}{2}} \scov \ncov^{-\frac{1}{2}}$ or equivalently $\mathcal{K} \propto \ncov^{-1} \scov \ncov^{-1}$.
To make $\ds$ directly comparable to \snr, we determine $\mathcal{K}$ and $\ds_0$ so that
\begin{equation}
    \ds = \frac{\vec{d}^\dagger \ncov^{-1} \scov \ncov^{-1} \vec{d} - \mathrm{Tr}\left[\ncov^{-1} \scov\right]}{\sqrt{\mathrm{Tr}\left[\left(\ncov^{-1} \scov\right)^2\right]}}, \label{eq:optimal}
\end{equation}
which satisfies $\left<\rho\right>|_{\epsilon=0}=0$ and $\mathrm{Var}[\ds]|_{\epsilon=0}=1$. 
We refer to this statistic as \coherent.

The same form of detection statistic was derived in the context of stochastic \gw signals in pulsar timing array experiments \cite{Anholm:2008wy,Ellis:2013nrb,Chamberlin:2014ria}.
In their studies, it was derived as a log likelihood ratio in the limit of a weak signal.
In fact, the log likelihood ratio in the limit of small $\epsilon$ is proportional to \coherent,
\begin{equation}
    \ln \frac{p(\vec{d}|\epsilon)}{p(\vec{d}|\epsilon=0)} = \epsilon^2 \sqrt{\mathrm{Tr}\left[\left(\ncov^{-1} \scov\right)^2\right]} \ds + \mathcal{O}(\epsilon^4).
\end{equation}

\subsection{Threshold and upper bounds} \label{sec:optimal_threshold}

The detection threshold for $\rho$, denoted by $\rho_{\rm th}$, is determined based on a pre-specified false alarm probability, $\fap$.
Let $p(\rho | \epsilon)$ represent the probability distribution of $\rho$ conditioned on the value of the coupling constant.
The relation between $\rho_{\rm th}$ and $\fap$ is given as follows:
\begin{equation}
    \fap \equiv \int_{\rho_{\rm th}}^{\infty}d\rho \, p(\rho |\epsilon = 0).
\end{equation}
On the other hand, given the observed value of $\rho$, denoted by $\rho_{\rm obs}$, the frequentist upper bound on $\epsilon$ with the confidence level of $\cl$, denoted by $\epsilon^{100\cl\,\%}$, is determined by
\begin{equation}
    1 - \cl = \int^{\rho_{\rm obs}}_0 d\rho \, p(\rho |\epsilon^{100\cl\,\%}).
\end{equation}

One efficient approach to solve those equations is to approximate the distribution as Gaussian distribution \cite{Chamberlin:2014ria},
\begin{equation}
    p(\rho |\epsilon) = \frac{1}{\sqrt{2 \pi \sigma^2}} \e^{-\frac{(\rho - \mu)^2}{2 \sigma^2}},
\end{equation}
where
\begin{align}
    &\mu = \epsilon^2 \sqrt{\mathrm{Tr}\left[\left(\ncov^{-1} \scov\right)^2\right]}, \\
    &\sigma^2 = \frac{\mathrm{Tr}\left[\left(\ncov^{-1} \scov \ncov^{-1} \dcov\right)^2\right]}{\mathrm{Tr}\left[\left(\ncov^{-1} \scov\right)^2\right]}.
\end{align}
However, the Gaussian approximation is not necessarily accurate enough and can significantly overestimate the significance of detection \cite{Hazboun:2023tiq}.

In the numerical simulations presented in Sec. \ref{sec:optimal_numerical}, we instead use a Monte-Carlo approach, similar to the method employed in \cite{KAGRA:2024ipf}.
For computing the threshold, we draw $\vec{d}$ from the Gaussian distribution with the covariance of $\ncov$, compute $\ds$ values with the simulated data, and compute their $100(1 - \fap)$--th percentile.
For computing the upper bounds, we draw $\vec{n}$ and $\vec{h}$ from the Gaussian distributions with the covariances of $\ncov$ and $\scov$ respectively, compute $\ds$ values with $\vec{d} = \vec{n} + \epsilon \vec{h}$ for varying values of $\epsilon$, and find out the value of $\epsilon$ that makes the $100(1 - \cl)$––th percentile of the $\ds$ values equal to $\ds_{\mathrm{obs}}$.


\subsection{Sensitivity} \label{sec:optimal_analytical}

In this section, we study how the sensitivity of $\rho$ scales with observation time and segment lengths.
Here we assume that the signal is detectable when $\left<\rho\right>=1$ and estimate the minimum value of $\epsilon$ that achieves it, as a measure of the sensitivity. 
We assume that the noise covariance matrix is diagonal, $\ncov_{fS, f'S'} = T_S P_{\mathrm{n}}(f, S) \delta_{ff'} \delta_{SS'}$.
We also assume that the noise \psd is almost constant in the frequency span where the signal is non-negligible and over the observation time, and it depends only on the detector index $I$, $P_{\mathrm{n}}(f, S) = P_{\mathrm{n}, I}$.
Then, $\left<\rho\right> = \epsilon^2 \sqrt{\mathrm{Tr}\left[\left(\ncov^{-1} \scov\right)^2\right]}$, where
\begin{equation}
    \begin{aligned}
        &\mathrm{Tr}\left[\left(\ncov^{-1} \scov\right)^2\right] = \\
        &~~\sum_{I,I'} \frac{1}{P_{\mathrm{n}, I} P_{\mathrm{n}, I'}} \sum_{s_I,s'_I} \frac{1}{T_S T_{S'}} \sum_{f,f'} \left| \scov_{fS, f'S'} \right|^2. \label{eq:trace}
    \end{aligned}
\end{equation}

The sum over $f$ and $f'$ is, in fact, taken over positive frequency indexes, $k$ and $k'$, where $f = k / T_{S}$ and $f' = k' / T_{S'}$.
Given $\scov_{(-f)S, (-f')S'} = \scov^\ast_{fS, f'S'}$, and assuming the correlation between positive and negative frequencies is negligible, we obtain
\begin{equation}
    \sum_{f, f'} \left| \scov_{fS, f'S'} \right|^2 = \frac{1}{2} \sum_{k = -\infty}^\infty \sum_{k' = -\infty}^\infty \left| \scov_{\frac{k}{T_S}S, \frac{k'}{T_{S'}}S'} \right|^2.
\end{equation}

Substituting Eq. \eqref{eq:hfS} into Eq. \eqref{eq:cov_def}, we obtain
\begin{equation}
    \begin{aligned}
        &\epsilon^2 \scov_{fS, f'S'} = \int^\infty_{-\infty} dt dt' \bigg[ \\
        &~~w(t - \bar{t}_S, S) w(t' - \bar{t}_{S'}, S') \left<h(t, I) h(t', I')\right> \times \\
        &~~\e^{-2 \pi \iu f (t - t_S) + 2 \pi \iu f' (t' - t_{S'})} \bigg].
    \end{aligned}
\end{equation}
By applying the Poisson summation formula,
\begin{equation}
    \sum_k \e^{2 \pi \iu \frac{k t}{T_S}} = T_S \sum_{n = -\infty}^\infty \delta\left(t - n T_S\right),
\end{equation}
we obtain
\begin{equation}
    \begin{aligned}
    &\frac{1}{T_S T_{S'}} \sum_{f,f'} \left| \scov_{fS, f'S'} \right|^2 = \frac{1}{2} \int^\infty_{-\infty} dt dt' \bigg[ \\
    &\left(w(t - \bar{t}_{\mathrm{S}}, S)\right)^2 \left(w(t' - \bar{t}_{\mathrm{S'}}, S')\right)^2 \left( R_{II'}(t, t') \right)^2 \bigg],
    \end{aligned}
\end{equation}
where $R_{II'}(t, t')$ is the correlation function of the signal with a unit coupling constant, $h(t, I; \epsilon=1)$, defined by
\begin{align}
    R_{II'}(t, t') &\equiv \frac{1}{\epsilon^2} \left<h(t, I) h(t', I')\right> \nonumber \\
    &= \left<h(t, I; \epsilon=1) h(t', I'; \epsilon=1)\right>.
\end{align}

Under the assumption that the window function for each segment is unity almost all over the segment, Eq. \eqref{eq:trace} is reduced to 
\begin{equation}
    \begin{aligned}
        &\mathrm{Tr}\left[\left(\ncov^{-1} \scov\right)^2\right] \simeq \\
        &\frac{1}{2} \sum_{I,I'} \frac{1}{P_{\mathrm{n}, I} P_{\mathrm{n}, I'}} \int_{t \in \mathcal{T}_I,t' \in \mathcal{T}_{I'}} dt dt' \left(R_{II'}(t, t')\right)^2, \label{eq:snr_3}
    \end{aligned}
\end{equation}
where $\mathcal{T}_I$ represents the observation time span of the $I$-th detector.
Hence, the minimum value of $\epsilon$ that achieves $\left<\rho\right>=1$ is given by
\begin{equation}
    \begin{aligned}
        &\epsilon_{\mathrm{th}} = \\
        &\left(\sum_{I,I'} \frac{1}{2 P_{\mathrm{n}, I} P_{\mathrm{n}, I'}} \int_{t \in \mathcal{T}_I,t' \in \mathcal{T}_{I'}} dt dt' \left(R_{II'}(t, t')\right)^2\right)^{-\frac{1}{4}}. \label{eq:epsilon_th}
    \end{aligned}
\end{equation}
Clearly, it does not depend on the choice of segment lengths in contrast with \incoherent, whose sensitivity gets worse as segments get shorter than $\tauc$.

To study how $\epsilon_{\mathrm{th}}$ scales with observation time, we assume that $R_{II'}(t, t';\mb)$ takes the following form,
\begin{equation}
    \begin{aligned}
       &R_{II'}(t, t') = \\
       &~~ D_I(t) D_{I'}(t') A(t' - t) \cos(\mb (t' - t) + \theta), \label{eq:rii_ansatz}
    \end{aligned}
\end{equation}
where $D_I(t)$ represents the detector response of the $I$--th detector and $A(t' - t) \cos(\mb (t' - t) + \theta)$ represents the auto-correlation function of an \ulbdm field or its spatial derivative.
As we have seen in Sec. \ref{sec:psd}, the non-oscillating part of the correlation function, $A(\tau)$, variates with the timescale of the signal coherence time $\tauc$ and vanishes for $|\tau| \gg \tauc$.
It can be easily seen that this functional form applies for the axion signal, the dark-photon signal in a KAGRA's auxiliary channel, and spin-2 \dm.
For the dilaton or dark-photon signal in a \gw channel, $R_{II'}(t, t')$ contains the auto-correlations and cross-correlations of the signals caused by finite light-traveling time and actual length changes, and each of these terms takes the functional form given in Eq. \eqref{eq:rii_ansatz}. 
The following arguments also apply to such cases.

In the following discussions, we assume that all the detectors are observing in a common time span, whose total duration is $T_{\mathrm{tot}}$, and $T_{\mathrm{tot}}$ is much larger than the variation time scale of $D_I(t)$.
This is a valid assumption for ground-based \gw detectors such as LIGO and KAGRA, where $D_I(t)$ varies on a timescale of a day, while $T_{\mathrm{tot}}$ typically spans months to years.
In addition, we assume that the variation time scale of $D_I(t)$ is much larger than $2 \pi / \mb$.
This assumption also holds for ground-based \gw detectors, as their target frequency is higher than $\mathcal{O}(1)\,\mathrm{Hz}$.

If $T_{\mathrm{tot}} \ll \tauc$, $A(|t' - t|) \simeq A(0)$ in the observation time span, and the integral over $t$ and $t'$ can be evaluated as follows,
\begin{equation}
    \int_{t \in \mathcal{T}_I,t' \in \mathcal{T}_{I'}} dt dt' \left(R_{II'}(t, t')\right)^2 \simeq \frac{1}{2} T_{\mathrm{tot}}^2 \left(A(0)\right)^2 \bar{D^2_I} \bar{D^2_{I'}},
\end{equation}
where
\begin{equation}
    \bar{D^2_I} \equiv \lim_{T \to \infty} \frac{1}{T} \int^{\frac{T}{2}}_{-\frac{T}{2}} dt \left(D_{I}(t)\right)^2.
\end{equation}
Substituting it into Eq. \eqref{eq:epsilon_th}, we obtain
\begin{equation}
    \epsilon_{\mathrm{th}} \simeq \left(T_{\mathrm{tot}}\right)^{-\frac{1}{2}} \left(\left(A(0)\right)^2 \sum_{I, I'} \frac{\bar{D^2_I} \bar{D^2_{I'}}}{4 P_{\mathrm{n}, I} P_{\mathrm{n}, I'}}\right)^{-\frac{1}{4}}.
\end{equation}

On the other hand, if $T_{\mathrm{tot}} \gg \tauc$, the integral over $t$ and $t'$ can be approximated as follows,
\begin{align}
    &\int_{t \in \mathcal{T}_I,t' \in \mathcal{T}_{I'}} dt dt' \left(R_{II'}(t, t')\right)^2 \nonumber \\
    &\simeq \int_{t \in \mathcal{T}_I} dt \int^\infty_{-\infty} d \tau \left(R_{II'}(t, t + \tau)\right)^2 \nonumber  \\
    &\simeq \frac{1}{2} k_{II'} T_{\mathrm{tot}} \tauc \left(A(0)\right)^2 \bar{D^2_I} \bar{D^2_{I'}},
\end{align}
where $k_{II'}$ is an $\mathcal{O}(1)$ constant that depends on the functional forms of $A(\tau)$ and $D_I(t)$.
Substituting it into Eq. \eqref{eq:epsilon_th}, we obtain
\begin{equation}
    \epsilon_{\mathrm{th}} = \left(T_{\mathrm{tot}} \tauc\right)^{-\frac{1}{4}} \left(\left(A(0)\right)^2 \sum_{I, I'} \frac{k_{II'} \bar{D^2_I} \bar{D^2_{I'}}}{4 P_{\mathrm{n}, I} P_{\mathrm{n}, I'}}\right)^{-\frac{1}{4}}.
\end{equation}

In summary, $\epsilon_{\mathrm{th}}$ decreases as $T_{\mathrm{tot}}$ increases, with $\epsilon_{\mathrm{th}} \propto T_{\mathrm{tot}}^{-\frac{1}{2}}$ for $T_{\mathrm{tot}} \ll \tauc$ and $\epsilon_{\mathrm{th}} \propto \left(T_{\mathrm{tot}} \tauc\right)^{-\frac{1}{4}}$ for $T_{\mathrm{tot}} \gg \tauc$.
This scaling is achieved when signal coherence is maximally exploited, which shows that \coherent can coherently extract an \ulbdm signal from segments of arbitrary lengths.

\subsection{Numerical simulation} \label{sec:optimal_numerical}

We also conduct numerical simulations to evaluate the sensitivity of our detection statistic. 
Specifically, we simulate data for the KAGRA's MICH channel, which contains signals caused by a dark photon coupling to $B-L$.
To create these simulations, we generate $10^3$ signal realizations assuming a unit coupling constant $(\epsilon=1)$ and the same number of independent noise realizations.
We assume the SHM distribution for velocities and polarizations.
The signal is then scaled by $\epsilon$ and added to the noise to produce data realizations for varying values of $\epsilon$.
The noise is generated with the KAGRA's design sensitivity presented in \cite{Michimura:2020vxn} and the signal frequency $\fb$ is set to $100\,\mathrm{Hz}$.
The duration of the data is $10^5\,\mathrm{s}$, which is much longer than the signal coherence time, $\tauc \sim 10^4\,\mathrm{s}$.

\begin{figure}[t]
    \centering
    \includegraphics[width=0.45\textwidth]{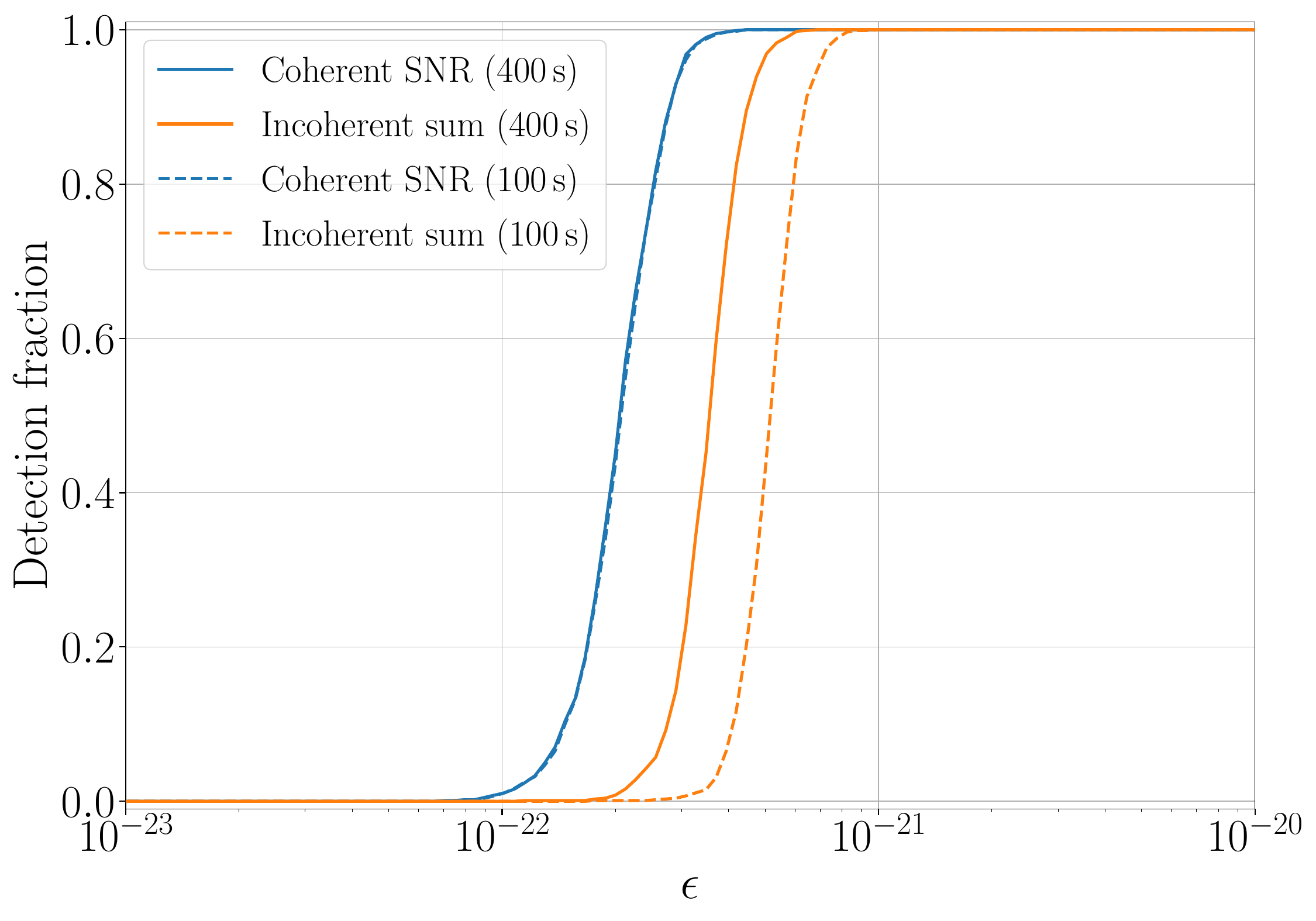}
    \caption{The fraction of data realizations in which \coherent (blue) or \incoherent (orange) exceeds the detection threshold at the false alarm probability of $10^{-4}$. Solid and dashed lines represent results for segment lengths of $400\,\mathrm{s}$ and $100\,\mathrm{s}$ respectively.}
    \label{fig:simulation_detection}
\end{figure}

For each realization, the data are divided into shorter segments and \coherent $\ds$ is computed.
We then evaluate the fraction of data realizations in which $\ds$ exceeds the detection threshold at the false alarm probability of $10^{-4}$.
Figure \ref{fig:simulation_detection} presents the detection fraction as a function of $\epsilon$, represented by the blue curves.
The simulations are performed with the two choices of segment duration, $T_S=400\,\mathrm{s}$ and $T_S=100\,\mathrm{s}$, whose results are presented as solid and dashed curves respectively.
When $\ds$ is computed, the signal correlation matrix is evaluated using Eqs. \eqref{eq:r_to_c} and \eqref{eq:analytic_r}, which are valid as the segment lengths are much shorter than $\tauc \sim 10^6 / \fb = 10^4\,\mathrm{s}$.
For comparison, the detection fraction obtained using \incoherent as a detection statistic is shown in orange.

As expected, the detection fraction increases with $\epsilon$, reflecting the increasing signal amplitude.
Notably, \coherent achieves a transition from 0 to 1 at smaller values of $\epsilon$, highlighting its superior sensitivity to \ulbdm.
Furthermore, the detection fraction for \coherent remains unaffected by the choice of $T_S$, while that for \incoherent decreases as $T_S$ gets smaller.
These results support the analytic arguments presented in the previous section.

The value of $\epsilon$ at which the detection fraction reaches $0.5$ is $2.1 \times 10^{-22}$ for \coherent, while the values for \incoherent are $3.6 \times 10^{-22}$ and $5.2 \times 10^{-22}$ with $T_S=400\,\mathrm{s}$ and $T_S=100\,\mathrm{s}$ respectively.
This indicates that \coherent achieves an improvement factor of $1.7$ for $T_S=400\,\mathrm{s}$ and $2.5$ for $T_S=100\,\mathrm{s}$.
These improvement factors roughly match the improvement factors expected by the difference in the coherence time of the analysis, $\left(\tauc / T_S\right)^{\frac{1}{4}} \simeq \left(10^4 / 400\right)^{\frac{1}{4}} \simeq 2.2$ and $\left(10^4 / 100\right)^{\frac{1}{4}} \simeq 3.2$.

\begin{figure}[t]
    \centering
    \includegraphics[width=0.45\textwidth]{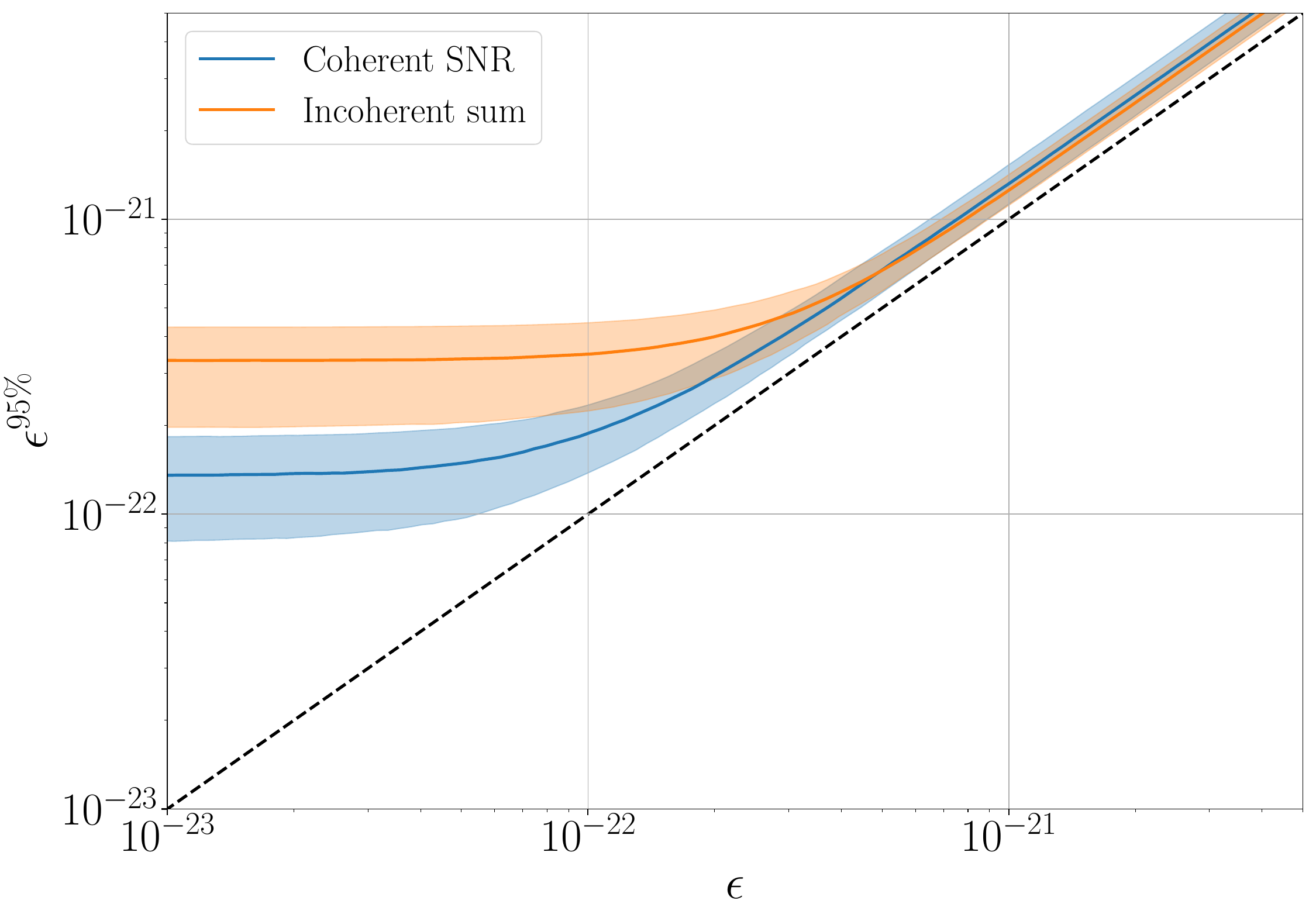}
    \caption{Upper bounds on $\epsilon$ computed with \coherent (blue) or \incoherent (orange) at the confidence level of $0.95$. The shaded regions represent the 1--$\sigma$ ranges of the bounds across all the data realizations, and the solid lines represent the medians.}
    \label{fig:simulation_upperbounds}
\end{figure}

We also compute the upper bounds on $\epsilon$ with $\beta = 0.95$ using both \coherent and \incoherent.
They are calculated with $T_S=100\,\mathrm{s}$.
The results are shown in Fig. \ref{fig:simulation_upperbounds}, where the shaded regions represent the 1--$\sigma$ ranges of the bounds across all realizations, and the solid lines represent the medians.
\Coherent provides tighter bounds for $\epsilon \lesssim 10^{-22}$, where signals are not confidently detected as shown in Fig. \ref{fig:simulation_detection}.
This demonstrates that \coherent provides better constraints by leveraging signal correlations even when the signal is not detected.
For higher values of $\epsilon$ where signals are confidently detected with either of the methods, the upper bounds become comparable.

\section{Bank of boson masses}  \label{sec:template_bank}

In real experiments, the mass of the \ulbdm is not known a priori.
Thus, the search is conducted over a grid of $\mb$, where the grid resolution must be fine enough to ensure that any potential signal is not missed.
This challenge is analogous to constructing a template bank in \gw searches that rely on known waveforms \cite{Sathyaprakash:1991mt,Owen:1995tm}.
For this reason, we refer to the grid as a ``mass bank".
In this section, we derive the requirements for the mass bank to ensure that the loss in \coherent due to its incompleteness remains negligible.


Suppose that data contain an \ulbdm signal with the boson mass of $\mb$ while \coherent is computed assuming the boson mass of $\mb'$. 
The expectation value of \coherent, $\left<\rho\right>$, is maximized when $\mb' = \mb$ while it diminishes when there is a mismatch, $\mb' \neq \mb$.
The fractional loss in $\left<\rho\right>$ due to this mismatch is given by
\begin{equation}
    \begin{aligned}
        &L(\mb, \mb') = \\
        &1 - \frac{\mathrm{Tr}\left[\ncov^{-1} \scov(\mb) \ncov^{-1} \scov(\mb')\right]}{\sqrt{\mathrm{Tr}\left[\left(\ncov^{-1} \scov(\mb)\right)^2\right]} \sqrt{\mathrm{Tr}\left[\left(\ncov^{-1} \scov(\mb')\right)^2\right]}},
    \end{aligned}
\end{equation}
where we explicitly show the dependence of the signal correlation matrix $\scov$ on the boson mass $\mb$.
To ensure that the fractional loss is negligible for an arbitrary value of $\mb$, $L(\mb, \mb') \ll 1$ should be satisfied for any neighboring masses in the mass bank.

By performing calculations similar to those done in Sec. \ref{sec:optimal_analytical}, we find that $L(\mb, \mb')$ can be written as follows,
\begin{widetext}
\begin{equation}
    \begin{aligned}
        &L(\mb, \mb') \simeq \\
        &~~1 - \frac{\sum_{I,I'} \frac{1}{P_{\mathrm{n}, I} P_{\mathrm{n}, I'}} \int_{t \in \mathcal{T}_I,t' \in \mathcal{T}_{I'}} dt dt' R_{II'}(t, t'; \mb) R_{II'}(t, t'; \mb')}{\sqrt{\sum_{I,I'} \frac{1}{P_{\mathrm{n}, I} P_{\mathrm{n}, I'}} \int_{t \in \mathcal{T}_I,t' \in \mathcal{T}_{I'}} dt dt' \left(R_{II'}(t, t'; \mb)\right)^2} \sqrt{\sum_{I,I'} \frac{1}{P_{\mathrm{n}, I} P_{\mathrm{n}, I'}} \int_{t \in \mathcal{T}_I,t' \in \mathcal{T}_{I'}} dt dt' \left(R_{II'}(t, t'; \mb')\right)^2}}, \label{eq:fractional_loss}
    \end{aligned}
\end{equation}
\end{widetext}
where we write the correlation function as $R_{II'}(t, t';\mb)$ to explicitly show its dependence on $\mb$.
Consequently, the condition $L(\mb, \mb') \ll 1$ holds if the overlap integral between $R_{II'}(t, t'; \mb)$ and $R_{II'}(t, t'; \mb')$ is close to its value at $\mb = \mb'$.

We again assume that all the detectors are observing in a common time span, whose total observation time is $T_{\mathrm{tot}}$ and also assume the functional form given in Eq. \eqref{eq:rii_ansatz} for $R_{II'}(t, t';\mb)$.
With the new integration variable, $\tau = t' - t$, the overlap integral can be expressed as follows,
\begin{widetext}
    \begin{equation}
        \begin{aligned}
            &\int_{t \in \mathcal{T}_I,t' \in \mathcal{T}_{I'}} dt dt' R_{II'}(t, t'; \mb) R_{II'}(t, t'; \mb') \\
            &= \int_{t \in \mathcal{T}_I,t' \in \mathcal{T}_{I'}} dt d \tau \left(D_I(t)\right)^2 \left(D_{I'}(t + \tau)\right)^2 A(\tau; \mb) A(\tau; \mb') \cos \left(\mb \tau + \theta (\mb) \right) \cos \left(\mb' \tau + \theta (\mb') \right),
        \end{aligned}
    \end{equation}
\end{widetext}
where the dependence of $A(\tau)$ and $\theta$ on $\mb$ is explicitly shown.
We then approximate $A(\tau;\mb)$ as constant for $|\tau| < \tauc$ and zero otherwise, that is, $A(\tau;\mb) = A(0;\mb) \Theta\left(\tauc - |\tau|\right)$.
With this approximation, we can easily see that the integral over $\tau$ starts to cancel when $|\mb' - \mb|$ is on the order of $2 \pi / \tauc$ if $\Ttot > \tauc$, and $2 \pi / \Ttot$ if $\Ttot < \tauc$.

Therefore, the condition $L(\mb, \mb') \ll 1$ is satisfied when the following inequality holds,
\begin{equation}
    |\mb' - \mb| \ll \frac{2 \pi}{\mathrm{min} \left[2 \pi / (\mb \vvir^2), T_{\mathrm{tot}}\right]}. \label{eq:bank_requirement}
\end{equation}
A sufficiently fine mass bank can be constructed by placing grid points of $\mb$ such that the difference between neighboring points satisfies Eq. \eqref{eq:bank_requirement}.

\section{Conclusion}  \label{sec:conclusion}

In this paper, we introduce a novel detection statistic, \coherent, specifically designed for the \ulbdm signal.
Unlike the \incoherent method employed in our previous studies, \coherent incorporates signal correlations across different data segments, enabling coherent integration over these segments.
In Sec. \ref{sec:optimal}, we demonstrate, through analytical arguments and numerical simulations, that the sensitivity of \coherent is independent of segment lengths and that it can detect weaker signals compared to \incoherent.
Additionally, to make this method applicable to an \ulbdm signal with an unknown mass, we discuss the required resolution of the mass grid to ensure no potential signal is missed in Sec. \ref{sec:template_bank}.
We argue that the grid spacing should satisfy $|\mb' - \mb| \ll 2 \pi / \mathrm{min} \left[2 \pi / (\mb \vvir^2), T_{\mathrm{tot}}\right]$ to achieve this goal.


In this work, we assume that instrumental noise can be modeled as a stationary Gaussian random process.
However, in real experiments, non-Gaussian and non-stationary noise can significantly amplify the detection statistic, leading to numerous false detections.
To improve robustness against such strong noise disturbances, the method proposed in \cite{Miller:2020vsl} counts the number of segments where the power exceeds a predefined threshold, rather than summing the power across all segments.
Furthermore, the signal correlation calculated in our study can be useful for distinguishing between noise artifacts and an \ulbdm signal.
Enhancing the robustness of the search while fully exploiting the signal coherence is an important direction for future work.

Beyond differentiating between noise artifacts and a signal, some authors have explored methods to distinguish various types of \dm by analyzing correlations between detectors and the shape of power spectra \cite{Miller:2022wxu,Manita:2023mnc}.
The signal correlation presented here can also enhance distinguishing capabilities in these contexts.


\section*{Acknowledgement}
S. M. acknowledges support from JSPS Grant-in-Aid for Transformative Research Areas (A) No.~23H04891 and No.~23H04893. 
J. K. acknowledges support from the JSPS Overseas Research Fellowships and from Istituto Nazionale di Fisica Nucleare (INFN) through the Theoretical Astroparticle Physics (TAsP) project.
Y. M. acknowledges support from JSPS KAKENHI Grant Nos.~20H05850, 20H05854, 20H05639, and 24K00640.

This material is based upon work supported by NSF's LIGO Laboratory which is a major facility fully funded by the National Science Foundation. This research has made use of data or software obtained from the Gravitational Wave Open Science Center (gwosc.org), a service of the LIGO Scientific Collaboration, the Virgo Collaboration, and KAGRA. This material is based upon work supported by NSF's LIGO Laboratory which is a major facility fully funded by the National Science Foundation, as well as the Science and Technology Facilities Council (STFC) of the United Kingdom, the Max-Planck-Society (MPS), and the State of Niedersachsen/Germany for support of the construction of Advanced LIGO and construction and operation of the GEO600 detector. Additional support for Advanced LIGO was provided by the Australian Research Council. Virgo is funded, through the European Gravitational Observatory (EGO), by the French Centre National de Recherche Scientifique (CNRS), the Italian Istituto Nazionale di Fisica Nucleare (INFN) and the Dutch Nikhef, with contributions by institutions from Belgium, Germany, Greece, Hungary, Ireland, Japan, Monaco, Poland, Portugal, Spain. KAGRA is supported by Ministry of Education, Culture, Sports, Science and Technology (MEXT), Japan Society for the Promotion of Science (JSPS) in Japan; National Research Foundation (NRF) and Ministry of Science and ICT (MSIT) in Korea; Academia Sinica (AS) and National Science and Technology Council (NSTC) in Taiwan.



\bibliographystyle{apsrev4-1}
\bibliography{reference}

\end{document}